\documentclass[aps,prd,superscriptaddress,showpacs,showkeys,a4paper,superscriptaddress,10pt,notitlepage]{revtex4-1}

\usepackage[utf8]{inputenc}
\usepackage[english]{babel}
\usepackage{amsmath}
\usepackage{amssymb}
\usepackage{amsfonts}
\usepackage{graphicx}
\usepackage{bm}
\usepackage{cancel}         
\usepackage{upgreek}          
\usepackage{mathtools}             
\usepackage[T1]{fontenc}

\usepackage{xcolor}
\definecolor{link_blue}{RGB}{52,46,157}

\usepackage[colorlinks,citecolor=link_blue,linkcolor=link_blue,urlcolor=link_blue]{hyperref}
\usepackage[tiny,raggedright]{titlesec}             
\usepackage[a4paper,textwidth=440pt,textheight=640pt,top=100pt,left=80pt]{geometry}

\setcitestyle{super}

\renewcommand{\vec}{\boldsymbol}

\newcommand{\Ei}{\mathrm{Ei}}
\newcommand\ri{\mathrm{i}}

\DeclareMathOperator{\re}{Re}

\allowdisplaybreaks[2]

\begin{document}
	
\title{A generating integral for the matrix elements of the Coulomb
  Green's function with the Coulomb wave functions}

\author{K. Dzikowski}

\affiliation{Max Planck Institute for Nuclear Physics, Saupfercheckweg
  1, 69117 Heidelberg, Germany}

\author{O. D. Skoromnik}
\email[Corresponding author: ]{olegskor@gmail.com}                

\affiliation{Max Planck Institute for Nuclear Physics, Saupfercheckweg
  1, 69117 Heidelberg, Germany}

\begin{abstract}
  We analytically evaluate the generating integral
  $K_{nl}(\beta,\beta') = \int_{0}^{\infty}\int_{0}^{\infty} e^{-\beta
    r - \beta' r'}G_{nl}(r,r') r^{q} r'^{q'} dr dr'$ and integral moments
  $J_{nl}(\beta, r) = \int_{0}^{\infty} dr' G_{nl}(r,r') r'^{q}
  e^{-\beta r'}$ for the reduced Coulomb Green's function
  $G_{nl}(r,r')$ for all values of the parameters $q$ and $q'$, when
  the integrals are convergent. These results can be used in
  second-order perturbation theory to analytically obtain the complete
  energy spectra and local physical characteristics such as electronic
  densities of multi-electron atoms or ions.
\end{abstract}



\maketitle

\section{Introduction}
\label{sec:introduction}

It has recently been demonstrated \cite{skoromnik_analytic_2017} that
a multi-electron atom can be effectively described via a simple model
using an effective charge. In this approach one starts from a
Hamiltonian of a multi-electron atom in the secondary-quantized
representation, written in the hydrogen-like basis with an effective
charge $Z^{*}$ being a free parameter. The effective charge depends
only on the set of occupation numbers of a given state and a charge of
an atom. Then one constructs perturbation theory
\cite{skoromnik_analytic_2017, hameka_erratum:_1968, *hameka_use_1967,
  PhysRev.178.126, hostler_corrections_1981, Adkins_2018,
  adkins_hydrogen_2018} (PT), where corrections to energy levels and
wave functions are given in terms of the matrix elements of a reduced
Coulomb Green's function (RCGF) $G_{nl}(\vec r,\vec r')$ with the
hydrogen like wave functions.

Due to the nature of the Coulomb field the radial and angular
variables are separated, which allows one to integrate out the angular
parts in matrix elements. The remaining radial integrals are further
reduced to the computation of the generating integral from RCGF of a
form
\begin{equation}
  K_{nl}(\beta,\beta') = \int_0^\infty \int_{0}^{\infty}
  e^{-\beta r -\beta' r'} G_{nl}(r,r')r^q{r'}^{q'} dr dr', \label{eq:1}
\end{equation}
which is convergent for $q \geq 0$ and $q' \geq 0$.  In
Eq.~(\ref{eq:1}) the letter $l$ denotes the index of angular
expansion.

In principle, one can use the direct numerical integration of
Eq.~(\ref{eq:1}). However, if the matrix elements can be evaluated
analytically then the model will provide analytical expressions for
the observable characteristics of multi-electron atoms with
Hartree-Fock accuracy \cite{skoromnik_analytic_2017}. Moreover, when
the indices $n$ and $l$ in Eq.~(\ref{eq:1}) are large, which is the
typical situation for Rydberg states \cite{sibalic_rydberg_2018} the
direct numerical integration of Eq.~(\ref{eq:1}) becomes very
inefficient or even impossible due to the increasing number of nodes
of the integrand. Furthermore, if repeated evaluations of the
generating integrals are required the efficient scheme of computation
of Eq.~(\ref{eq:1}) is desirable.  Consequently, in our work we
analytically evaluate the generating integral $K_{nl}(\beta, \beta')$
and integral moments
\begin{align}
  J_{nl}(\beta, r) = \int_{0}^{\infty} G_{nl}(r,r') r'^{q}
  e^{-\beta r'} dr' \label{eq:2}
\end{align}
for all values of the parameters $q$ and $q'$ when the integrals are
convergent.

A starting point for the evaluation of (\ref{eq:1}) and (\ref{eq:2})
is the work of Johnson and Hirschfelder \cite{johnson_radial_1979},
who derived the explicit expressions for the RCGF in terms of
elementary functions and computed integral moments
\begin{align}
  \int_{0}^{\infty} dr' G_{nl}(r,r') r'^{k+2} e^{-Z r'/n} \label{eq:3}
\end{align}
of the RCGF, where $k\geq -l - 2$. This expression is a special case
of Eq.~(\ref{eq:2}) when $\beta = Z/n$.

In addition, we mention the work of Hill and Huxtable
\cite{hill_generating_1982}, who evaluated the generating integral
(\ref{eq:1}) for the case of $q,q' = l+1$ and provided recurrence
relations, which allow one to compute the generating integral with the
increasing powers of $r,r'$ starting from $q,q' = l+1$. Their result
is based on treating the generating integral as a Laplace transform of
$\beta$ and $\beta'$ and solving the differential equation for this
Laplace transform. However, the generating integral in the range
$0 \leq q,q' \leq l$ has not been evaluated. Consequently, here we
evaluate the generating integral and integral moments for this
case. In addition, we provide a closed form result for values
$q,q' > l+1$, which is directly applicable without the use of
recurrence relations. Finally, we extend $J_{nl}(\beta,r)$ to values
of $\beta$ when the integral is convergent.

In our work we follow the approach of the direct integration of the
RCGF with the corresponding powers of $r,r'$ and exponentials. The
main complications come from the fact, that when $0 \leq q,q' < l+1$,
the individual terms of the RCGF possess singularities that are
explicitly cancelled only upon summing all expressions together.

The article is organized in the following way. For the readers
convenience in Sec.~\ref{sec:result} we summarize the main results
without any derivations. In Sec.~\ref{sec:comp-analyt-eval} we compare
the evaluation time based on our analytical expressions with the
direct numerical evaluation of the integrals. In
Secs.~\ref{sec:derivation}-\ref{sec:proofs} we provide the details of
all derivations. Finally, a \texttt{Mathematica} notebook has been
prepared as a supplementary information, where we have programmed the
main results of the article.


\subsection{Definitions}
\label{sec:definitions}

The Hydrogen-like wave functions $\psi(\vec r)$ satisfy the
Schr\"{o}dinger equation
\begin{align}
  \left[-\frac{1}{2}\nabla^{2} - \frac{Z}{r} -
  E\right]\psi(\vec r) = 0. \label{eq:4}
\end{align}
Here $E$ is the energy of the system, $Z$ is the charge of the nucleus
and the atomic units are employed.

The variables in Eq.~(\ref{eq:4}) are separated in spherical
coordinates \cite{LandauQM}. Consequently, the eigenfunctions of a
discrete spectrum are represented as a product
\begin{align}
  \psi_{nlm}(\vec r) = R_{nl}(r)Y_{lm}(\Omega), \label{eq:5}
\end{align}
where
\begin{align}
  R_{nlZ}(r)
  &= \sqrt{Z^{3}\frac{(n-l-1)!}{(n+l)!}}
    \frac{2}{n^{2}} t^{l} e^{-t/2} L_{n-l-1}^{2l+1}(t), \label{eq:6}
  \\
  t
  &= \frac{2Zr}{n}, \nonumber
\end{align}
with $L_{n-l-1}^{2l+1}(t)$ being the associated Laguerre
polynomials~\cite{gradstejn_table_2009} and $Y_{lm}(\Omega)$ the
spherical harmonics~\cite{LandauQM}. An analogous expression can be
obtained for the continuous spectrum as well \cite{LandauQM}.

The associated Laguerre polynomials can be expanded according to their
definition, thus leading to
\begin{align}
  R_{nlZ} (r) = Z^{3/2} R_{nl}(r Z)
  &= Z^{3/2} e^{-\frac{Z r}{n}} \sum_{i=l+1}^{n} N_{n,l}(i)
    (rZ)^{i-1}, \label{eq:7}
  \\
  N_{n,l}(i)
  &= \sqrt{\frac{(n-l-1)!}{(n+l)!}} \frac{1}{n}
    \left(\frac{-2}{n}\right)^i
    \frac{(-1)^{l+1}}{(i-1-l)!}{{n+l}\choose{n-i}}, \label{eq:8}
\end{align}
where the binomial is defined as ${{n}\choose{k}} = n!/(k!(n-k)!)$.

In Eqs.~(\ref{eq:4})--(\ref{eq:8}) $n$ is an integer, $n>0$, $l$ is
an integer, $0 \leq l \leq n-1$ and $m$ is an integer, $-l \leq m \leq
l$.

The bound states of Eq.~(\ref{eq:4}) are described by energy
eigenvalues $E_{n} = -Z^{2}/(2n^{2})$.

The Green's function of the Hydrogen like atom satisfies the equation
\cite{johnson_radial_1979}
\begin{align}
  \left[-\frac{1}{2}\nabla^{2} - \frac{Z}{r} -
  E\right]G(\vec r, \vec r'; E) = -\delta(\vec r - \vec r') \label{eq:9}
\end{align}
and can be expanded over the eigenfunctions of Eq.~(\ref{eq:4})
\begin{align}
  G(\vec r, \vec r'; E)
  &= \sideset{}{^*}\sum_{n' = 1}^{\infty} \sum_{l = 0}^{n'-1}
    \sum_{m = -l}^{l} \frac{\psi_{n'lm}(\vec r)\psi^{*}_{n'lm}(\vec
    r')}{E - E_{n'}} \label{eq:10}
  \\
  &= \sum_{l = 0}^{\infty} \sum_{m = -l}^{l}
    Y_{lm}(\Omega)Y^{*}_{lm}(\Omega') \sideset{}{'}\sum_{n' = l+1}^{\infty}
    \frac{R_{n'l}(r)R^{*}_{n'l}(r')}{E - E_{n'}} = \sum_{l = 0}^{\infty}
    \sum_{m = -l}^{l} Y_{lm}(\Omega)Y^{*}_{lm}(\Omega') G_{l}(r,r';
    E). \nonumber
\end{align}
Here the sum with asterisk denotes the summation over both the discrete and
continuous spectra.

When $E$ equals the energy of the bound state the CGF has a pole.

The radial CGF $G_{l}(r,r';E)$ satisfies the equation
\cite{johnson_radial_1979}
\begin{align}
  (H_{l} - E) G_{l}(r,r';E) = \left[-\frac{1}{2r}
  \frac{\partial^{2}}{\partial r^{2}}r + \frac{l(l+1)}{2r^{2}} -
  \frac{Z}{r} - E\right] G_{l}(r,r';E) =
  -\frac{\delta(r - r')}{rr'} \label{eq:11}
\end{align}
and is expressed in terms of the Whittaker functions
\cite{abramowitz_handbook_2013}
\begin{align}
  G_{l}(r,r';E)
  &= -\frac{4Z}{\nu}
  \frac{\Gamma(l+1-\nu)}{t_{<}t_{>}}M_{\nu, l+1/2}(t_{<}) W_{\nu,
  l+1/2}(t_{>}), \label{eq:12}
  \\
  t_{<}
  &= \min(t,t'), \quad t_{>} = \max(t, t'), \quad \nu =
  \sqrt{-\frac{Z^{2}}{2E}}, \nonumber
\end{align}
with $\Gamma(t)$ being the gamma function \cite{gradstejn_table_2009}.

The RCGF is defined as a CGF from which a state with the principal
quantum number $n$ is subtracted \cite{PhysRev.178.126}
\begin{align}
   G_{n}(\vec r, \vec r')
  &= \sideset{}{^*}\sum_{\substack{n' = 1 \\ n' \neq n}}^{\infty}
  \sum_{l = 0}^{n'-1}
    \sum_{m = -l}^{l} \frac{\psi_{n'lm}(\vec r)\psi^{*}_{n'lm}(\vec
    r')}{E_{n} - E_{n'}} \label{eq:13}
  \\
  &= \sum_{l = 0}^{\infty} \sum_{m = -l}^{l}
    Y_{lm}(\Omega)Y^{*}_{lm}(\Omega')
    \sideset{}{^*}\sum_{\substack{n' = l+1 \\ n' \neq n}}^{\infty}
    \frac{R_{n'l}(r)R^{*}_{n'l}(r')}{E_{n} - E_{n'}} = \sum_{l = 0}^{\infty}
    \sum_{m = -l}^{l} Y_{lm}(\Omega)Y^{*}_{lm}(\Omega')
  G_{nl}(r,r'). \nonumber
\end{align}

The summation in RCGF $G_{nl}(r,r')$ starts from $l+1$. Therefore, one
needs to distinguish the two distinct cases. When $l \leq n - 1$ the
term with $n' = n$ must be explicitly excluded. For the opposite
situation of $l \geq n$ no such term appears anyway, thus making the
restriction $n \neq n'$ redundant.

As shown in \cite{johnson_radial_1979}, for the case of $l \geq n$ the
RCGF, satisfying the equation
\begin{align}
  (H_{l} - E_{n}) G_{nl}(r,r') = -\frac{\delta(r - r')}{rr'}, \label{eq:14}
\end{align}
is given by ($t = 2Zr/n$, $t' = 2Zr'/n$):
\begin{align}
  G_{nl}(r,r')
  &= (-1)^{l+1-n} \frac{4Z}{n} (l-n)! (l+n)! (tt')^{-l-1}
  e^{-(t+t')/2} \sum_{i = 0}^{l+n} {{2l-i}\choose{l-n}}
  \frac{t_{>}^{i}}{i!} \nonumber
  \\
  &\times \left[e^{t_{<}} \sum_{j = 0}^{l-n} {{2l-j}\choose{l+n}}
    \frac{(-t_{<})^{j}}{j!} - \sum_{k = 0}^{l+n} {{2l-k}\choose{l-n}}
    \frac{t_{<}^{k}}{k!}\right], \label{eq:15}
\end{align}
while for $l \leq n-1$, the RCGF satisfying an equation
\begin{align}
  (H_{l} - E_{n}) G_{nl}(r,r') = R_{nl}(r') R_{nl}(r) -\frac{\delta(r
  - r')}{rr'}. \label{eq:16}
\end{align}
reads:
\begin{align}
  G_{nl}(r,r')
  &= \frac{4Z}{n} \frac{(n-l-1)!}{(n+l)!} e^{-(t+t')/2}
  \Bigg\{ L_{n-l-1}^{2l+1}(t) L_{n-l-1}^{2l+1}(t') \nonumber
  \\
  &\times \left[\ln{t} + \ln{t'} + \frac{t+t'}{2n} - \psi(n-l)
  -\psi(n+l+1) - \frac{4l + 5}{2n} - \Ei(t_{<})\right] \nonumber
  \\
  &+L_{n-l-1}^{2l+1}(t)\left[\frac{t'}{n} L_{n-l-2}^{2l+2}(t')
    + \sum_{k = 0}^{n-l-2}A_{k} t^{\prime k}\right] +
    L_{n-l-1}^{2l+1}(t')\left[\frac{t}{n} L_{n-l-2}^{2l+2}(t)
  + \sum_{k = 0}^{n-l-2}A_{k} t^{k}\right] \nonumber
  \\
  &+L_{n-l-1}^{2l+1}(t_{>}) e^{t_{<}} \Phi_{nl}(t_{<}) -
  L_{n-l-1}^{2l+1}(t_{<})\sum_{k = 1}^{2l+1} {{n+l}\choose{n-l-1+k}}
    \frac{(k-1)!}{t_{>}^{k}} \nonumber
  \\
  &+ L_{n-l-1}^{2l+1}(t_{>})\sum_{k = 1}^{2l+1} 
    \frac{(k-1)!}{t_{<}^{k}} \left[{{n+l-k}\choose{n-l-1}} e^{t_{<}} -
    {{n+l}\choose{n-l-1+k}}\right]\Bigg\}, \label{eq:17}
\end{align}
where
\begin{align}
  A_{k}
  &= \frac{(-1)^{k}}{k!}{{n+l}\choose{n-l-1-k}}\sum_{j =
    k+1}^{n-l-1}\frac{2j + 2l+1}{j(j+2l+1)}, \label{eq:18}
  \\
  \Phi_{nl}(x)
  &= \sum_{j =
    1}^{n-l-1}\frac{(-x)^{j}}{j!} {{n+l}\choose{n-l-1-j}} \sum_{k =
    1}^{j} \frac{(k-1)!}{x^{k}}. \label{eq:19}
\end{align}

\subsection{The main result}
\label{sec:result}

\subsubsection{Generating integrals}
\label{sec:generating-integrals}

When $l +1 \le n$ the closed form main result is given via\footnote{We
  work in the rescaled variables, i.e., $x = Zr$, $x' = Zr'$ and
  $\lambda = \beta / Z$, $\lambda' = \beta' / Z$. Therefore, the
  factor from the Jacobian $Z^{-q-q'-2}$ should be included when
  relating $K_{nl}(\beta, \beta')$ and $K_{nl}(\lambda, \lambda')$.}
\begin{align}
  &K_{nl}(\lambda, \lambda') = \int e^{-\lambda x - \lambda' x'}
    G_{nl}(x,x')x^{q} x'^{q'} dx dx' \label{eq:20}
  \\
  &=2Z \sum_{i_1=l}^{n-1}
    \sum_{i_2=1-l}^{1+l} {{n-l-1}\choose{n-i_1-1}} (-1)^{i_1+l+1}
    \frac{(i_2-1+l)!}{(i_1+l+1)!}
    \Big(\frac{2}{n}\Big)^{i_1-i_2+1} \nonumber
  \\
  &\mspace{50mu} \times \Bigg\{ {{n-i_2}\choose{n-l-1}}
    \left[f_{q-i_2+1}^{q'+i_1+1} \left(\alpha',\alpha-\frac{2}{n}\right)
    + f_{q'-i_2+1}^{q+i_1+1} \left(\alpha, \alpha'- \frac{2}{n} \right)
    \right] \nonumber
  \\
  &\mspace{190mu} - {{n+l}\choose{n-1+i_2}}
    \left[{g}_{q-i_2+1}^{q'+i_1+1}(\alpha,\alpha') +
    {g}_{q'-i_2+1}^{q+i_1+1}(\alpha',\alpha) \right]\Bigg\} \nonumber
  \\
  &\mspace{30mu}+2Z\sum_{i_1=l+1}^{n} \sum_{i_2=l+1}^{n}
    {{n-l-1}\choose{n-i_1}}
    {{n+l}\choose{n-i_2}}
    \frac{(-2/n)^{i_1+i_2-1}}{(i_1+l)!(i_2-l-1)!}\nonumber
  \\
  &\mspace{70mu}\times\Bigg\{\frac{(q+i_1-1)!}{\alpha^{q+i_1}}
  \frac{(q'+i_2-1)!}{\alpha^{q'+i_2}} \Bigg(\Psi(q+i_1) + \Psi(q'+i_2)
    -\ln\left(\frac{n^2}{4}\alpha \alpha'\right) +A_{i_1,i_2} \nonumber
  \\
  &\mspace{250mu}+\frac{(2n+l-i_1+1)(q+i_1)}{(i_1+l+1)\alpha n^2} +
  \frac{(2n+l-i_2+1)(q+i_2)}{(i_2+l+1)\alpha' n^2} \Bigg)
  \nonumber
  \\ 
  &\mspace{70mu}
   - \left(f_{q+i_1}^{q'+i_2}\left(\alpha',\alpha-\frac{2}{n}\right)
    +
    f_{q'+i_1}^{q+i_2}\left(\alpha,\alpha'-\frac{2}{n}\right)\right)
    B_{i_2}+ \left(\frac{n}{2}\right)^{q+q'+i_1+i_2}
    I_{q+i_1-1}^{q'+i_2-1}\left(\frac{n}{2}\alpha, \frac{n}{2}\alpha'\right)
    \Bigg\}, \nonumber
\end{align}				
where $\alpha = \lambda+\frac{1}{n}$ and
$\alpha' = \lambda'+\frac{1}{n}$ and the constants $A_{i_{1},i_{2}}$
and $B_{i}$ are given by:
\begin{align}
  A_{i_1,i_2}
  &= -\frac{4l+5}{2n} + \Psi(n+l+1) + \Psi(n-l) - \Psi(i_1-l)
  \nonumber
  \\
  &\mspace{30mu}- \Psi(1+i_1+l) - \Psi(i_2-l) -
    \Psi(1+i_2+l), \label{eq:21}
  \\
  B_{i}
  &=\frac{(n-i)}{(i+l+1)(i-l)} {_3F_2}(1,1,1-n+i,1+i-l,2+i+l,1) \nonumber
  \\
  &=\sum_{k=1}^{n-i}
    \frac{(i-l-1)!(i+l)!(n-i)!(-1)^{k}(k-1)!}
    {(k+i-l-1)!(k+i+l)!(n-i-k)!}, \label{eq:22}
\end{align}
where $\Psi(n)$ is the logarithmic derivative of the Gamma function
\cite{gradstejn_table_2009} (also known as the zeroth polygamma
function) and $_pF_q(a_{1},\ldots,a_{p};b_{1},\ldots,b_{q};z)$ is the
generalized hypergeometric function, which in this case can be
evaluated as a finite series (see Eq.~(\ref{eq:22})).

The function $f_{a}^{b}(x,y)$ is defined as follows:
\begin{subequations}
\begin{itemize}
\item $a>0$ and $a+b>0$
  \begin{align}
    f_{a}^{b} (x,y) = \frac{\Gamma(a+b)}{a x^b}{
    _2F_1}\left(a,a+b,a+1,-\frac{y}{x}\right), \label{eq:23}
  \end{align}
\item $a = -n$ is a non-positive integer and $a+b > 0$:
  \begin{align}
    f_{-n}^{b} (x,y)
    &= 2\frac{\Gamma(b)}{x^{b}} \frac{(-y)^n}{n!}
      \left(\Psi(b)-\ln{x}\right) + \sum_{i=0}^{n-1}
    \frac{\Gamma(b-n+i)}{(i-n)x^{b-n+i}} \frac{(-y)^i}{i!} \nonumber
    \\
    &+ b! \frac{(-y)^{1+n}}{x^{b+1}} 
      {_3\tilde{F}_2}\left(1,1,b+1,n+2,2,-\frac{y}{x}\right), \label{eq:24}
  \end{align}
\item $a+b = -m$ is a non-positive integer and $a > 0$
  \begin{align}
    f_{a}^{-a-m} (x,y)
    &= \sum_{i=0}^m
      \frac{(-x)^{m-i}}{(m-i)!(i+a)} \frac{(-y)^i}{i!} \left(\Psi(m+1-i) - \ln{x} - \frac{1}{2(i+a)} \right)
      \nonumber
    \\
    &+(a+m)!\frac{(-y)^{m+1}}{x}
      {_3\tilde{F}_2}(1,1,a+m+1,m+2,2+a+m,-\frac{y}{x}), \label{eq:25}
  \end{align}
\item $a+b = -m$ is a non-positive integer and $b > 0$
  \begin{align}
    f_{-m-b}^{b} (x,y)
    &= \sum_{i=0}^m
      \frac{(-x)^{m-i}}{(m-i)!(i-m-b)} \frac{(-y)^i}{i!}
      \left(\Psi(m+1-i) - \ln{x} - \frac{1}{2(i-m-b)} \right)
      \nonumber
    \\
    &+\sum_{i=1+m}^{m+b-1} \frac{\Gamma(i-m)}{(i-m-b)x^{i-m}}
      \frac{(-y)^i}{i!} + 2\frac{\Gamma(b)}{x^{b}}
      \frac{(-y)^{m+b}}{(m+b)!} (\Psi(b) - \ln{x}) \nonumber
    \\
    &+ b!\frac{(-y)^{m+b+1}}{x^{b+1}}
      {_3\tilde{F}_2}
      \left(1,1,1+b,2,2+m+b,-\frac{y}{x}\right), \label{eq:26}
  \end{align}
\item $a = -m$ and $b = -n$ are both non-positive integers
  \begin{align}
    f_{-m}^{-n} (x,y)
    &= \sum_{i=0,i \neq m}^{m+n}
      \frac{(-x)^{m+n-i}}{(m+n-i)!(i-m)} \frac{(-y)^i}{i!}
      \left(\Psi(m+n+1-i) - \ln(x) - \frac{1}{2(i-m)}\right) \nonumber
    \\
    &+ \frac{(-x)^n}{n!} \frac{(-y)^{m}}{m!}
      \left((\Psi(n+1)-\ln(x))^2-\Psi^{(1)}(n+1) + \frac{\pi^2}{3}\right) \nonumber
    \\
    &+ n!\frac{(-y)^{m+n+1}}{x}
      {_3\tilde{F}_2}
      \left(1,1,1+n,m+n+2,n+2,-\frac{y}{x}\right). \label{eq:27}
  \end{align}
\end{itemize}
\end{subequations}
When $y=0$ the function $f_{a}^{b}(x,0)$ becomes:
\begin{subequations}
\begin{itemize}
\item  $a+b>0$ and $a \neq 0$:
 \begin{align}
    f_{a}^{b} (x,0) = \frac{\Gamma(a+b)}{a x^{a+b}}, \label{eq:28}
  \end{align}
\item $b > 0$ and $a = 0$
 \begin{align}
    f_{0}^{b} (x,0) = \frac{2\Gamma(b)}{x^b}(\Psi(b)-\ln x). \label{eq:29}
  \end{align}
\item $a+b = -m$ is a non-positive integer and $a \neq 0$:
 \begin{align}
    f_{a}^{-a-m} (x,0) = \frac{(-x)^m}{a m!}
   \left(\Psi(1+m)-\ln(x)-\frac{1}{2 a}\right), \label{eq:30}
  \end{align}
\item $b = -m$ is a negative integer and $a = 0$ 
 \begin{align}
    f_{0}^{-m} (x,0) =
   \frac{(-x)^m}{m!} \left(\frac{\pi^2}{3} + [\Psi(1+m)-\ln(x)]^2 -
   \Psi^{(1)}(1+m)\right). \label{eq:31}
  \end{align}
\end{itemize}
\end{subequations}

The function ${g}_{a}^{b}(x,y)$ is defined as follows:
\begin{subequations}
\begin{itemize}
\item $a>0$ and $b>0$
  \begin{align}
    {g}_{a}^{b} (x,y) =
    \frac{\Gamma(a)}{x^a}\frac{\Gamma(b)}{y^b}, \label{eq:32}
  \end{align}
\item $a = -n$ is a non-positive integer
  \begin{align}
    {g}_{-n}^{b} (x,y) =
    \frac{\Gamma(b)}{y^b}\frac{(-x)^n}{n!}(\Psi(1+n)+\Psi(b)-\ln(x
    y)), \label{eq:33}
  \end{align}
\end{itemize}
\end{subequations}
The function $I_{m}^{n}$ is defined as:
\begin{subequations}
\begin{align}
  I_{m}^{n}(x,y) = \frac{n!}{{y}^{n+1}}
  \sum_{i=0}^{n}\frac{{y}^i}{i!} f_{m+1}^i \left(y-1, x\right) +
  \frac{m!}{x^{m+1}} f_{n+1}^0(-1, y), \label{eq:34}
\end{align}
and when $y = 1$
\begin{align}
  I_{m}^{n}(x,1) = -n!
  \left(\sum_{t=1}^{n}\frac{1}{t!}f_{t}^{m+1}(x,0)+\frac{m!}{x^{m+1}}(H_{m} -
  \ln(x))\right). \label{eq:35}
\end{align}
\end{subequations}

In addition, in Eqs.~(\ref{eq:20})--(\ref{eq:33})
${_{3}\tilde{F}_{2}(a,b,c;d,e;z)}$ is the generalized regularized
hypergeometric function \cite{gradstejn_table_2009} and in
Eq.~(\ref{eq:35}) $H_{n}$ is a harmonic number:
$H_n = 1+\frac{1}{2}+\frac{1}{3}+...+\frac{1}{n}$.

The closed form result for $l \geq n$ reads
\begin{align}
  K_{nl}(\lambda, \lambda')
  &= \int e^{-\lambda x - \lambda' x'} G_{nl}(x,x')x^{q} x'^{q'} dx dx'
    \nonumber
  \\
  &=(-1)^{n+l} Z n \sum_{i=-l}^n \sum_{j=-l}^n \frac{(l-i)!}{(l+j)!}
    {{l+n}\choose{i+l}} \left(\frac{2}{n}\right)^{i+j} \nonumber
  \\
  &\mspace{90mu}\times \Bigg\{{{l-j}\choose{l+n}}(-1)^{j+l}
    \left[f_{j+q}^{i+q'}\left(\alpha',\alpha -\frac{2}{n}\right) +
    f_{j+q'}^{i+q}\left(\alpha,\alpha' - \frac{2}{n} \right)\right]
    \nonumber
  \\
  &\mspace{130mu}-{{l-j}\choose{l-n}}
    \left[f_{j+q}^{i+q'}\Big(\alpha',\alpha\Big)+f_{j+q'}^{i+q}
    \Big(\alpha,\alpha'\Big)\right] \Bigg\}. \label{eq:36}
\end{align}

\subsubsection{Integral moments}
\label{sec:integral-moments}

The main result for the integral moments for the case $n \ge l+1$
reads ($x = Zr$, $\lambda = \beta / Z$)
\begin{align}
  J_{nl}(\lambda, x) =
  &Z^{-q-1}\int e^{-\lambda y} G_{n,l}(x,y) {y}^{q} dy  \nonumber
  \\
  = &2Z^{-q} e^{-\frac{x}{n}}\Bigg\{\sum_{i_1=l}^{n-1} \sum_{i_2=1-l}^{1+l}
    {{n-l-1}\choose{n-i_1-1}} (-1)^{i_1+l+1} \frac{(i_2-1+l)!}{(i_1+l+1)!}
    \left(\frac{2}{n}\right)^{i_1-i_2+1} \nonumber
  \\
  &\mspace{60mu}\times \Bigg[{{n-i_2}\choose{n-l-1}}
    \left(x^{i_1}\mathcal{F}_{q+1-i_2}(2/n-\alpha,x) +
    \frac{x^{-i_2} e^{2x/n}}{\alpha^{1+i_1+q}}\Gamma(1+i_1+q,x
    \alpha)\right) \nonumber
  \\
  &\mspace{100mu}-{{n+l}\choose{n-1+i_2}} \left(x^{i_1}\mathcal{G}_{q+1-i_2}
    (\alpha) + \frac{(q+i_1)!}{x^{i_2}\alpha^{1+q+i_1}}
    \right) \Bigg] \nonumber
  \\
  &\mspace{40mu}+\sum_{i_1=l+1}^{n} \sum_{i_2=l+1}^{n} {{n-l-1}\choose{n-i_1}}
    \frac{(-2/n)^{i_1+i_2-1}}{(i_1+l)!(i_2-l-1)!}\nonumber
  \\
  &\mspace{90mu}\times\Bigg[\left( x^{i_1-1}
    \mathcal{F}_{q+i_2}\left(\frac{2}{n}-\alpha,x\right) + \frac{x^{i_2-1}}{\alpha^{i_1+q}}e^{\frac{x}{n}}
    \Gamma(i_1+q,x \alpha) \right) B_{i_2} \nonumber
  \\
  &\mspace{120mu}-x^{i_1-1} \frac{(q+i_2-1)!}{\alpha^{q+i_2}}
    \Bigg(\ln\left(\frac{4}{n^2}\frac{x}{\alpha}\right) + \Psi(q+i_2) +
    \frac{2n+l-i_1+1}{i_1+l+1}\frac{x}{n^2} \nonumber
  \\
  &\mspace{300mu}+ \frac{2n+l-i_2+1}{i_2+l+1}\frac{i_2+q}{n^2
    \alpha}+A_{i_1,i_2}\Bigg) \nonumber
  \\
  &\mspace{120mu} -x^{i_1-1}\left(\frac{n}{2}\right)^{q+i_2}
    \mathcal{I}_{q+i_2-1}(\alpha n/2,2x/n)\Bigg]\Bigg\}, \label{eq:37}
\end{align}
where the function $\mathcal{F}$ is defined as:
\begin{subequations}
\begin{align}
  \mathcal{F}_{q}(x,y) = \frac{y^q}{q} {_1F_1}(q,q+1,x y) \label{eq:38}
\end{align}
and in the case of $q$ being a non-positive integer:
\begin{align}
  \mathcal{F}_{-n}(x,y)
  &= \sum_{i=0}^{n-1} \frac{x^{i}}{i-n}\frac{y^{i-n}}{i!}
    +\frac{x^n}{n!} \ln y + \frac{y x^{1+n}}{(1+n)!}
    {_2F_2}(1,1,2,2+n,xy), \label{eq:39}
\end{align}
while in the case of $x=0$:
\begin{align}
  \mathcal{F}_{q}(0,y) &= \frac{y^q}{q}\mspace{120mu}\mathcal{F}_{0}(0,y) = \ln y. \label{eq:40}
\end{align}
\end{subequations}

In addition, the function $\mathcal{I}$ is defined as:
\begin{subequations}
\begin{align}
  \mathcal{I}_{q}(x,y) = \frac{q!}{x^{q+1}} \left(\Ei(y-x y) -
  \ln(1-x) + \sum_{i=1}^q \frac{(x y)^i}{i i!} {_1F_1}(i,i+1,y-y
  x)\right) \label{eq:41}
\end{align}
and for $x = 1$
\begin{align}
  \mathcal{I}_{q}(1,y) =
  q!\left(\ln y + \gamma+\sum_{i=1}^q\frac{y^i}{i
  i!}\right), \label{eq:42}
\end{align}
\end{subequations}
while the function $\mathcal{G}$ is:
\begin{subequations}
\begin{align}
  \mathcal{G}_{q}(x) = \frac{\Gamma(q)}{x^{q}} \label{eq:43}
\end{align}
and for negative integers:
\begin{align}
  \mathcal{G}_{-n}(x) =
  \frac{(-x)^n}{n!}(\Psi(1+n)-\ln(x)). \label{eq:44}
\end{align}
\end{subequations}

The closed form result for the case $n \le l$ is given via ($x = Zr$,
$\lambda = \beta / Z$)
\begin{align}
  J_{nl}(\lambda, x) =
  &Z^{-q-1}\int e^{-\lambda y}G_{nl}(x,y){y}^{q}dy \nonumber
  \\
  &=(-1)^{n+l}Z^{-q}n e^{-\frac{x}{n}} \sum_{i=-l}^n \sum_{j=-l}^n
     \frac{(l-i)!}{(l+j)!}{{n+l}\choose{i+l}}
     \Big(\frac{2}{n}\Big)^{i+j} \nonumber
  \\
  &\mspace{20mu}\times \Bigg[{{l-j}\choose{l+n}} (-1)^{j+l}
    \left(x^{i-1}\mathcal{F}_{j+q}(2/n-\alpha,x) +
    \frac{x^{j-1}}{\alpha^{i+q}} e^{2x/n} \Gamma(i+q,x \alpha)\right)
    \nonumber
  \\
  &\mspace{60mu}-{{l-j}\choose{l-n}}
    \left(x^{i-1}\mathcal{F}_{j+q}(-\alpha,x) +
    \frac{x^{j-1}}{\alpha^{i+q}} \Gamma(i+q,x \alpha)\right)
    \Bigg]. \label{eq:45}
\end{align}

\subsection{Comparison of evaluation time of an analytical calculation with the numerical one}
\label{sec:comp-analyt-eval}

In this section we compare the evaluation time of the generating
integral via analytical expressions~(\ref{eq:20}) and~(\ref{eq:51})
with the direct numerical evaluation of the integral when the term
with $E_{n}=-Z^{2}/(2n^{2})$ has been explicitly subtracted from the
Green's function Eq.~\eqref{eq:12} (See also supplementary
information). Since this expression is divergent, we introduce a
parameter $\delta$ and substitute the energy as $E_{n} - \ri \delta$,
effectively regularizing the numerical integration. That is, we
evaluate the integrals from the expression
$G_{l}(r,r'; E_{n} - \ri \delta) - R_{nl}(r)R_{nl}^{*}(r')/(-\ri
\delta)$.  Since the resulting answer for the generating integral is
$\delta$ dependent we always ensure that the numerical value is
independent of the $\delta$ by varying the parameter $\delta$. For all
comparisons we employed the standard desktop Intel 2600k 3.4GHz
processor \cite{intel_2600k}.

\begin{table}[b]
  \begin{tabular}{| c | c | c | c | c | c | c |}
    \hline 
    $n$ & $l$ & $q$ & $q'$ & Analytic time & Numeric time & $\delta$\\
    \hline
    3 & 1 & 2 & 0 & 8$\times 10^{-3}$ s & 3.83 s & $10^{-4}$\\
    \hline
    7 & 5 & 4 & 1 & 0.128 s & 2.39 s & $10^{-4}$\\
    \hline
    16 & 15 & 10 & 10 & 0.188 s & NaN & NaN\\
    \hline 
    37 & 1 & 1 & 0 & 8.38 s & 39.6 s & $10^{-7}$\\
    \hline
  \end{tabular}
  \caption{Comparison of computational times of the generating
    integral for different values of the parameters $n$, $l$, $q$ and
    $q'$ of analytical expressions (\ref{eq:20}) and (\ref{eq:83})
    with the direct numerical integration using Mathematica
    \cite{Mathematica}, when the energy is shifted by $\ri\delta$ from
    its resonant value. The last column shows the value of $\delta$
    required to obtain the accurate results to four significant
    figures. NaN means that the integral did not converge to the
    correct value for any values of $\delta$.}\label{tab:1}
\end{table}

\begin{table}
  \begin{tabular}{| c | c | c | c | c | c | c |}
    \hline 
    $n$ & $l$ & $q$ & $\lambda$ & Analytic time & Numeric time & $\delta$
    \\
    \hline
    1 & 0 & 0 & 0.37 & 0.19 s & 21.7 s & $10^{-4}$
    \\
    \hline
    2 & 4 & 3 & 0.37 & 0.01 s & 32.2 s & $10^{-4}$
    \\
\hline
    5 & 4 & 7 & 0.37 & 3.89 s & 68.1 s & $10^{-5}$
    \\
    \hline
    6 & 8 & 8 & 0.37 & 0.05 s & 53.5 s & $10^{-6}$
    \\
    \hline  \end{tabular}
  \caption{Comparison of computational times of the integral moments
    for different values of parameters $n$, $l$, $q$ and $\lambda$ of
    analytical expressions with the direct numerical integration using
    Mathematica \cite{Mathematica} at 100 different values of $x$.
    The last column shows the value of $\delta$ required to obtain the
    accurate results to four significant figures.}\label{tab:2}
\end{table}

\begin{table}
  \begin{tabular}{|c|c|c|c|c|}
    \hline 
    Element & Atomic number & Analytic time & Numeric time & $\delta$\\
    \hline
    Li & 3 & 3.26 s & 29.3 s & $10^{-3}$\\
    \hline
    F & 9 & 14.2 s & 261 s & $10^{-4}$\\
    \hline
    Ne & 10 &  29.3 s & 321 s & $10^{-4}$\\
    \hline 
  \end{tabular}
  \caption{Comparison of computational times of second order
    single-electron correction to ground state energies of some
    example neutral atoms.} \label{tab:3}
\end{table}

In Tab.~\ref{tab:1} we provide the summary of the evaluation times of
the generating integral for different values of parameters $n$, $l$,
$q$ and $q'$. As can be observed from the table the numerical
evaluation is several orders of magnitude slower then our analytical
expressions. As follows from Tab.~\ref{tab:1} the evaluation time
increases significantly when $n \gg l$. Consequently, the evaluation
time of some high-$n$ Rydberg states is still large and requires
further optimization. For example, our analytic results allow one to
derive asymptotic expressions for large values of $n$.

In Tab.~\ref{tab:2} we compare the evaluation time of analytically
computed integral moments with the numerically computed ones for some
values of the parameters $q$, $n$, $l$ and $\lambda$. In this
simulation, we numerically evaluated the integral moments at one
hundred different values of $r$ in order to compare with an
analytically calculated curve. As in the situation of the generating
integral the evaluation time of the direct numerical integration is a
few orders of magnitude slower.

Moreover, as was explained in the introduction the generating integral
can be used to calculate the second order single-electron corrections
to energies of neutral atoms and ions in the effective charge model of
a multi-electron atom \cite{skoromnik_analytic_2017}. Therefore, we
compare a few cases of evaluation times between our analytical and
numerical approaches, which is given in Tab.~\ref{tab:3}. Typically a
fully sequential version of an \texttt{effz} program from
Ref.~\cite{skoromnik_analytic_2017} implemented in Mathematica
requires one order of magnitude larger times for the direct numerical
evaluation as compared to the analytical expressions.

As a final test (see supplementary information), we evaluated
generating integrals for the large number of input parameters, which
corresponds to the repeated evaluation of integrals. We considered 81
examples of $K_{nl}(\lambda, \lambda')$ and $J_{nl}(\lambda, x)$ each
($n \in \langle2,4 \rangle$, $l \in \langle0,2 \rangle$,
$q \in \langle0,2 \rangle$, $\lambda \in \langle1,2 \rangle$) for a
total compuaion time of 1.43 s and 0.192 s respectively, as compared
to 348 s and 4.62 s numerically.

Evaluation time of numerical integrals of the RCGF depends primarily
on the principal quantum number $n$ due to the oscilatory behaviour of
Rydberg states \cite{sibalic_rydberg_2018}. This happens due to the
increasing number of nodes of the integrand, resulting in oscillatory
behavior. Consequently, the accurate evaluation of the integral
demands the smaller and smaller values of $\delta$ to keep the
constant accuracy, since in many cases large positive values are
almost completely cancelled by large negative values. Therefore, the
precise result would require a forbiddingly accurate evaluation of the
integrand at every point.

On the other hand, evaluation time of our analytical results is harder
to investigate as it depends strongly on the methods used for
computation of hypergeometric functions, appearing in the main
results. This is further complicated by the fact that hypergeometric
functions with integer coefficients can be in most cases expressed by
elementary functions. This suggests that optimization of associated
algorithms could reduce the evaluation time of the analytical
calculaion, which is however beyond the scope of this
paper. Nevertheless, we have found out that in all relevant cases the
total evaluation time through analytical expressions is of the order
of 0.001-0.1 seconds on Intel 2600k 3.4GHz processor.

\section{Derivation of the main results}
\label{sec:derivation}

We want to evaluate the following integrals:
\begin{align}\label{eq:46}
  K_{nl}(\lambda, \lambda')
  &= \int e^{-\lambda x - \lambda' x'} G_{nl}(x,x')x^{q} x'^{q'} dx dx',
  \\
  J_{nl}(\lambda, x)
  &= \int e^{-\lambda y}G_{nl}(x,y){y}^{q}dy. \label{eq:47}
\end{align}

The main difficulty arises from the fact that the integrals of the
individual terms of the RCGF contain divergences in the case when $q$
or $q'$ are integers smaller than $l$ and these divergences cancel out
upon summing all terms together. For this reason, the strategy
employed in our work is to first derive expressions valid for
non-integer values of $q$ and $q'$, then find the Laurent series of
\eqref{eq:46} and \eqref{eq:47} around $q = m + \delta$,
$q' = m' + \delta$, where $m$ and $m'$ are non-negative integers and
finally show that for $q,q'\geq0$ the divergent parts (terms
proportional to $\delta^{-1}$ and $\delta^{-2}$) always vanish.

\subsection{Evaluation of the auxiliary integrals}
\label{sec:eval-auxil-integr}

First, we evaluate the following auxiliary integral
\begin{align}
  \int_0^\infty \int_r^\infty e^{-\lambda r - \lambda' r'} r^{a-1}
  {r'}^{b-1} dr' dr
  &= \int_0^\infty e^{-\lambda r} r^{a-1}
  {\lambda'}^{-b} \Gamma(b, \lambda' r)dr \nonumber
  \\
  &= \frac{\Gamma(a+b)}{a{\lambda'}^{a+b}}
    {_2F_1}\left(a, a+b, a + 1,
    -\frac{\lambda}{\lambda'}\right), \label{eq:48}
\end{align}
convergent whenever $\re a>0$, $\re(a + b) >0$ and
$\re(\lambda + \lambda') > 0$. Here ${_2F_1}(a,b,c,z)$ is the Gauss
Hypergeometric function \cite{gradstejn_table_2009}. Let us introduce
the following notation
\begin{align}
  u_{a}^{b}(x,y) = \frac{\Gamma(a+b)}{ax^{a+b}}{
  _2F_1}\left(a, a+b, a+1, -\frac{y}{x}\right) =
  \sum_{i=0}^\infty \frac{\Gamma(a+b+i)}{(a+i)x^{a+b+i}}
  \frac{(-y)^i}{i!}. \label{eq:49}
\end{align}

In the case when $a$ and $b$ are positive integers the expression
(\ref{eq:49}) simplifies if one uses the contingous relations
\cite{gradstejn_table_2009} for ${_2F_1}(a,b,c,z)$. This allows us to
rewrite it as a finite sum
\begin{align}
  u_{a}^{b} (x,y) = \frac{\Gamma(a)}{y^a} \left(\frac{\Gamma(b)}{x^b} -
  \left(x+y\right)^{-b} \sum_{i=0}^{a-1} \frac{\Gamma(b+i)}{i!}
  \left(\frac{y}{y+x}\right)^i\right). \label{eq:50}
\end{align}

Furthermore, for $b=0$ and $a$ a positive integer we get:
\begin{align}
  u_{a}^{0} (x,y) = \frac{\Gamma(a)}{y^a} \left(\ln{\left(1 +
  \frac{y}{x}\right)} - \sum_{i=1}^{a-1} \frac{1}{i}
  \left(\frac{y}{y+x}\right)^i \right). \label{eq:51}
\end{align}

Eq.~\eqref{eq:49} is useful when working with modern computer algebra
software such as Mathematica \cite{Mathematica} that can automatically
simplify ${_2F_1}(a,b,c,z)$ for given integer parameters, while
\eqref{eq:50} and \eqref{eq:51} are useful for evaluating $u_a^b$ with
finite precision arithmetic, as they require evaluating an explicitly
finite amount of terms.

Finally, the introduced functions $u_{a}^{b}$ allow us to evaluate
the following integral:
\begin{align}
  &\int_0^\infty e^{\mu r_{<}-\lambda r - \lambda' r'} r^q
    {r'}^{q'} r_{<}^{p} r_{>}^{p'} dr dr' \nonumber
  \\
  &= \int_0^\infty \int_r^\infty e^{(\mu-\lambda) r - \lambda' r'}
    r^{q+p} {r'}^{q'+p'} dr' dr + \int_0^\infty \int_{r'}^\infty
    e^{(\mu-\lambda') r' - \lambda r} r^{q+p'} {r'}^{q'+p} dr dr'
    \nonumber
  \\
  &=
    u_{q+p+1}^{q'+p'+1}(\lambda',\lambda-\mu) +
    u_{q'+p+1}^{q+p'+1}(\lambda,\lambda'-\mu). \label{eq:52}
\end{align}

\subsection{Derivation for $l \geq n$}
\label{sec:derivation-l-geq}

In this case the derivation is straightforward. We employ the
definition of the RCGF (\ref{eq:15}) and shift the index of summation
to get:
\begin{align}
  G_{nl}(x,y)
  &=(-1)^{n+l}n \sum_{i=-l}^n \sum_{j=-l}^n \frac{(l-i)!}{(l+j)!}
    {{n+l}\choose{i+l}} \left(\frac{2}{n}\right)^{i+j} e^\frac{-x-y}{n}
    \max[x,y]^{i-1} \min[x,y]^{j-1} \nonumber
  \\
  &\times \left[{{l-j}\choose{l+n}} (-1)^{j+l} e^{2/n \min[x,y]}
          - {{l-j}\choose{l-n}} \right], \label{eq:53}
\end{align}
Then we simply use Eq.~\eqref{eq:52} to write the answer in terms of
$u^a_b(x,y)$:
\begin{align}
&\int e^{-\lambda x - \lambda' y} G_{nl}(x,y)x^{q} y^{q'} dx dy
    =(-1)^{n+l} Z n \sum_{i=-l}^n \sum_{j=-l}^n \frac{(l-i)!}{(l+j)!}
    {{l+n}\choose{i+l}} \left(\frac{2}{n}\right)^{i+j} \nonumber
  \\
  &\mspace{90mu}\times \Bigg\{{{l-j}\choose{l+n}}(-1)^{j+l}
    \left[u_{j+q}^{i+q'}\left(\alpha',\alpha -\frac{2}{n}\right) +
    u_{j+q'}^{i+q}\left(\alpha,\alpha' - \frac{2}{n} \right)\right]
    \nonumber
  \\
  &\mspace{130mu}-{{l-j}\choose{l-n}}
    \left[u_{j+q}^{i+q'}\Big(\alpha',\alpha\Big)+u_{j+q'}^{i+q}
    \Big(\alpha,\alpha'\Big)\right] \Bigg\}. \label{nl}
\end{align}

\subsection{Derivation for $n>l$}
\label{sec:derivation-nl}

In the case of $n>l$, we split the Green's function into two parts:
\begin{equation}
  G_{nl}(x,y) = \frac{4Z}{n} \frac{(n-l-1)!}{(n+l)!}
  \left(G_{nl}^{(\mathrm{sg})}(x,y) +
  G_{nl}^{(\mathrm{nsg})}(x,y)\right), \label{eq:54}
\end{equation}
where $G^{(\mathrm{sg})}_{nl}$ contains all terms that are singular
around the origin and $G^{(\mathrm{nsg})}_{nl}$ those that are not. In
terms of explicit powers of $x$ and $y$, we then get:
\begin{align}
  G_{nl}^{(\mathrm{sg})}(x,y)
  &= e^{\frac{-x-y}{n}} \sum_{i_1=l}^{n-1} \sum_{i_2=1-l}^{1+l}
    {{n+l_1}\choose{n-i_1-1}} (-1)^{i_1+l} \frac{(i_2-1+l)!}{(i_1-l)!}
    \left(\frac{2}{n}\right)^{i_1-i_2}\nonumber
  \\
  &\times \left[{{n-i_2}\choose{n-l-1}} e^{2/n\min[x,y]}
    \frac{\max[x,y]^{i_1}}{\min[x,y]^{i_2}} - {{n+l}\choose{n-1+i_2}}
    \Big(\frac{x^{i_1}}{y^{i_2}} + \frac{y^{i_1}}{x^{i_2}} \Big)
    \right], \label{eq:55}
  \\
  G_{nl}^{(\mathrm{nsg})}(x,y)
  &= e^{\frac{-x-y}{n}} \sum_{i_1=l+1}^{n} \sum_{i_2=l+1}^{n}
    {{n+l}\choose{n-i_1}}{{n+l}\choose{n-i_2}}
    \frac{(-1)^{i_1+i_2+1}}{(i_1-l-1)!(i_2-l-1)!}
    \left(\frac{2}{n}\right)^{i_1+i_2-2} \nonumber
  \\
  &\times \Bigg[\max[x,y]^{i_1-1}\min[x,y]^{i_2-1}e^{2/n\min[x,y]}
    B_{i_2} \nonumber
  \\
  &\mspace{40mu}- x^{i_1-1} y^{i_2-1} \Bigg(\ln\left(\frac{4}{n^2}x
    y\right) - \Ei\left(\frac{2}{n}\min[x,y]\right) + A_{i_1,i_2}
    \nonumber
  \\
  &\mspace{160mu}+ \frac{2n+l-i_1+1}{i_1+l+1}
    \frac{x}{n^2} + \frac{2n+l-i_2+1}{i_2+l+1}
    \frac{y}{n^2}\Bigg)\Bigg], \label{eq:56}
\end{align}
where $\Ei(x)$ is the exponential integral function
\cite{gradstejn_table_2009} and the constants $A_{i_{1},i_{2}}$ and
$B_{i}$ are defined in Eqs.~(\ref{eq:21})-(\ref{eq:22}).
			
We can use Eq.~\eqref{eq:52} to immediately evaluate the elements of
$G^{(\mathrm{sg})}_{nl}$, as:
\begin{align}
  &\int  e^{-\lambda x - \lambda' y}G_{nl}^{(\mathrm{sg})}(x,y)
    x^{q}y^{q'} dx dy \nonumber
  \\
  &= \sum_{i_1=l}^{n-1} \sum_{i_2=1-l}^{1+l}
    {{n+l}\choose{n-i_1-1}} (-1)^{i_1+l} \frac{(i_2-1+l)!}{(i_1-l)!}
    \left(\frac{2}{n}\right)^{i_1-i_2} \nonumber
  \\
  &\mspace{60mu} \times \Bigg\{{{n-i_2}\choose{n-l-1}}
    \left[u_{q-i_2+1}^{q'+i_1+1} \left(\alpha',\alpha -
    \frac{2}{n}\right) +u_{q'-i_2+1}^{q+i_1+1} \left(\alpha,\alpha' -
    \frac{2}{n}\right)\right] \nonumber
  \\
  &\mspace{120mu}-{{n+l}\choose{n-1+i_2}}
    \left[\frac{(q+i_1)!}{\alpha^{1+q+i_1}}
    \frac{\Gamma(q'-i_2+1)}{{\alpha'}^{q'-i_2+1}} +
    \frac{\Gamma(q-i_2+1)}{\alpha^{q-i_2+1}} \frac{(q'+i_1)!}
    {{\alpha'}^{1+q'+i_1}} \right]\Bigg\}. \label{eq:57}
\end{align}
			
Before we can evaluate $G^{(\mathrm{nsg})}_{nl}$ we need expressions
for the terms containing logarithm and exponential integral
functions. The evaluation of the term containing the logarithmic
function is straightforward:
\begin{align}
  \int  e^{-\lambda x - \lambda' y} \ln(\mu x y) x^{q} y^{q'} dx dy =
  \frac{q!}{\lambda^{q+1}} \frac{q'!}{{\lambda'}^{q'+1}}
  \left(\Psi(q+1) + \Psi(q'+1) - \ln\left(\frac{\lambda
  \lambda'}{\mu}\right)\right), \label{eq:58}
\end{align}
but the exponential integral function is non-trivial:
\begin{align}
  I_{q,q'}(\lambda,\lambda') = \int e^{-\lambda x - \lambda' y} x^q
  y^{q'} \Ei\left(\min[x,y]\right) dx dy, \label{eq:59}
\end{align}
Therefore, we first evaluate a simple case:
\begin{align}
  I_{0,0}(\lambda,\lambda') = \int e^{-\lambda x - \lambda' y}
  \Ei\left(\min[x,y]\right) dx dy = -
  \frac{\ln(\lambda+\lambda'-1)}{\lambda \lambda'}, \label{eq:60}
\end{align}
and take derivatives, with respect to $\lambda$ and $\lambda'$:
\begin{align}
  I_{q,q'}(\lambda,\lambda')
  &=(-1)^{q+q'} \partial_\lambda^{(q)}\partial_{\lambda'}^{(q')}
    \frac{\ln(\lambda+\lambda'-1)}{\lambda \lambda'} \nonumber
  \\
  &=(-1)^{q+q'}\sum_{s}^q \sum_{t}^{q'} {{q}\choose{s}}
    {{q'}\choose{t}} \partial_\lambda^{(s)} \partial_{\lambda'}^{(t)}
    \ln(\lambda+\lambda'-1) \partial_\lambda^{(q-s)}
    \partial_{\lambda'}^{(q'-t)} \frac{1}{\lambda \lambda'} \nonumber
  \\
  &= (-1)^{q+q'}\sum_{s,t,s+t\neq 0}^{q, q'} {{q}\choose{s}}{{q'}\choose{t}}
    \frac{(-1)^{s+t+1}\Gamma(s+t)}{(\lambda+\lambda'-1)^{s+t}}
    \frac{(-1)^{s+q}(q-s)!}{\lambda^{q-s}}
    \frac{(-1)^{q'+t}(q'-t)!}{{\lambda'}^{q'-t}} \nonumber
  \\
  & \mspace{90mu}+\frac{(-1)^{q}(q)!}{\lambda^{q}}
    \frac{(-1)^{q'}(q')!}{{\lambda'}^{q'}} \log(\lambda+\lambda'-1)
    \nonumber
  \\
  &=-\frac{(q)!}{\lambda^{q}} \frac{(q')!}{{\lambda'}^{q'}}
    \Bigg(\sum_{s=0}^{q}\sum_{t=1}^{q'} \frac{\Gamma(s+t)}{s!t!}
    \frac{\lambda^s
    {\lambda'}^t}{(\lambda+\lambda'-1)^{s+t}} \label{eq:61}
  \\
  &\mspace{240mu}+ \sum_{s=1}^{q} \frac{\Gamma(s)}{s!}
  \frac{\lambda^s}{(\lambda+\lambda'-1)^{s}} -
  \ln(\lambda+\lambda'-1)\Bigg) \label{eq:62}
  \\
  &=\frac{q'!}{{\lambda'}^{q'+1}}
    \sum_{i=0}^{q'}\frac{{\lambda'}^i}{i!} u_{q+1}^i
    \left(\lambda'-1, \lambda\right) +
    \frac{q!}{\lambda^{q+1}} u_{q'+1}^0(-1,
    \lambda'), \label{eq:63}
\end{align}
where in the last step we have made use of Eqs.~\eqref{eq:50} and
\eqref{eq:51}. This gives us the integrals of the exponential integral
function as:
\begin{align}
  &\int  e^{-\lambda x - \lambda' y} \Ei\left(\mu \min[x,y]\right) x^q
    y^{q'} dx dy = \mu^{-2-q-q'}
    I_{q,q'}\left(\frac{\lambda}{\mu},\frac{\lambda'}{\mu} \right)
    \nonumber
  \\
  &= -\frac{q'!}{{\lambda'}^{q'+1}}
    \sum_{i=0}^{q'}\frac{{\lambda'}^i}{i!} u_{q+1}^i
    \left(\lambda'-\mu, \lambda\right) - \frac{q!}{\lambda^{q+1}}
    u_{q'+1}^0(-\mu, \lambda'). \label{eq:64}
\end{align}

The usage of contiguous relations for the hypergeometric function
introduced the indeterminacy into the expression (\ref{eq:64}) when
$\mu = \lambda'$. However, Eqs.~(\ref{eq:61})-(\ref{eq:62}) are well
defined. Therefore, we start from Eqs.~(\ref{eq:61})-(\ref{eq:62}) in
which we plug-in $\lambda / \lambda'$ for the first argument, one for
the second and apply the contiguous relations again. This yields
\begin{align}
  {\lambda'}^{-2-q-q'} I_{q,q'}\left(\frac{\lambda}{\lambda'},1\right)
  &= -\frac{(q)!}{\lambda^{q+1}} \frac{(q')!}{{\lambda'}^{q'+1}}
  \Bigg(\sum_{s=0}^{q}\sum_{t=1}^{q'} \frac{\Gamma(s+t)}{s!t!}
  \frac{{\lambda'}^{t}}{(\lambda)^{t}} + \sum_{s=1}^{q}
  \frac{\Gamma(s)}{s!}- \ln(\frac{\lambda}{\lambda'})\Bigg) \nonumber
  \\
  &=\frac{q'!}{{\lambda'}^{q'+1} \lambda^{q+1}} \left(\sum_{t=1}^{q'}
  \frac{1}{t!} u_{t}^{q+1} \left(\frac{\lambda}{\lambda'}, 0\right) +
  q! \left[H_q - \ln\left(\frac{\lambda}{\lambda'}\right) \right]
    \right). \label{eq:65}
\end{align}
Here $H_{q}$ is the harmonic number.

Finally, we evaluate elements of $G^{(\mathrm{nsg})}_{nl}$ that come
out as:
\begin{align}
  &\int e^{-\lambda x - \lambda' y} G_{nl}^{(\mathrm{nsg})}(x,y) x^{q} y^{q'} dx
    dy= \nonumber
  \\
  &\sum_{i_1=l+1}^{n} \sum_{i_2=l+1}^{n}
    {{n+l}\choose{n-i_1}}{{n+l}\choose{n-i_2}}
    \frac{(-1)^{i_1+i_2+1}}{(i_1-l-1)!(i_2-l-1)!}
    \left(\frac{2}{n}\right)^{i_1+i_2-2} \nonumber
  \\
  &\mspace{90mu}\times
    \Bigg\{\left(u_{q+i_1}^{q'+i_2}(\alpha',\alpha-\frac{2}{n}) +
    u_{q'+i_1}^{q+i_2}(\alpha,\alpha'-\frac{2}{n})\right) B_{i_2}
    \nonumber
  \\
  &\mspace{120mu}-\frac{(q+i_1-1)!}{\lambda^{q+i_1}}
    \frac{(q'+i_2-1)!}{\lambda^{q'+i_2}} \Bigg(\Psi(q+i_1) + \Psi(q'+i_2) -
    \ln\left(\frac{n^2}{4}\alpha \alpha'\right) +A_{i_1,i_2} \nonumber
  \\
  &\mspace{340mu}+\frac{(2n+l-i_1+1)(q+i_1)}{(i_1+l+1)\lambda n^2} +
    \frac{(2n+l-i_2+1)(q+i_2)}{(i_2+l+1)\lambda' n^2} \Bigg)
    \nonumber
  \\
  &\mspace{120mu} + \left(\frac{n}{2}\right)^{q+q'+i_1+i_2}
    I_{q+i_1-1,q'+i_2-1}\left(\frac{n}{2}\lambda,
    \frac{n}{2}\lambda'\right) \Bigg\},\label{eq:66}
\end{align}
finalizing the derivation. 

\section{Derivation of the integral moments}

We start from the evaluation of the following auxiliary integral
\begin{align}
  &\int^{\infty}_{0}e^{\mu \min[r,r']-\lambda r'}{r'}^q \min[r,r']^a
  \max[r, r']^b dr' \nonumber
  \\
  &=\frac{r^{1+a+b+q}}{1+a+q} {_1F_1}(1+a+q,2+a+q,r(\mu-\lambda)) +
    \frac{r^a}{\lambda^{1+b+q}} e^{\mu r} \Gamma(1+b+q,r \lambda) \nonumber
  \\
  &= r^{b} \mathcal{F}_{1+a+q}(\mu-\lambda,r) +
    \frac{r^a}{\lambda^{1+b+q}} e^{\mu r} \Gamma(1+b+q,r
    \lambda), \label{eq:67}
\end{align}
where for the purpose of dealing with singularities we define:
\begin{align}
  \mathcal{F}_{a}(x,y) = \frac{y^a}{a} {_1F_1}[a,a+1,x y] =
  \sum_{i=0}^\infty \frac{x^{i}}{a+i} \frac{y^{a+i}}{i!} \label{eq:68}
\end{align}
Integrating \eqref{eq:55}, gives:
\begin{align}
  &\int^{\infty}_{0} e^{-\lambda y} {y}^q G_{nl}^{(\mathrm{sg})}(x,
  y)dy \nonumber
  \\
  & \mspace{60mu}=e^{-\frac{x}{n}} \sum_{i_1=l}^{n-1} \sum_{i_2=1-l}^{1+l}
    {{n+l}\choose{n-i_1-1}} (-1)^{i_1+l} \frac{(i_2-1+l)!}{(i_1-l)!}
    \left(\frac{2}{n}\right)^{i_1-i_2} \nonumber
  \\
  &\mspace{90mu}\times \Bigg[{{n-i_2}\choose{n-l-1}} \left(x^{i_1}
    \mathcal{F}_{q+1-i_2}(\alpha-2/n,x) +
    \frac{x^{-i_2}}{\alpha^{1+i_1+q}} e^{2x/n} \Gamma(1+i_1+q,x
    \alpha)\right) \nonumber
  \\
  &\mspace{120mu}-{{n+l}\choose{n-1+i_2}}
    \left(\frac{x^{i_1}}{\alpha^{1+q-i_2}} \Gamma(q-i_2+1) +
    \frac{1}{x^{i_2} \alpha^{1+q+i_1}} \Gamma(q+i_1+1)
    \right) \Bigg]. \label{eq:69}
\end{align}
To integrate \eqref{eq:56} we first need:
\begin{equation}
  \int^{\infty}_{0} e^{-\lambda y} {y}^q \ln(x y)dy =
  \frac{q!}{\lambda^{q+1}} \left(\ln\left(\frac{x}{\lambda}\right) +
    \Psi(1+q)\right), \label{eq:70}
\end{equation}
as well as:
\begin{equation}
  \mathcal{I}_q(\lambda,x) = \int^{\infty}_{0} e^{-\lambda y} {y}^q
  \Ei[\mu \min(x,y)] dy, \label{eq:71}
\end{equation}
which is non-trivial. Therefore, we first evaluate Eq.~(\ref{eq:71})
when $q = 0$, $\mu = 1$
\begin{equation}
  \mathcal{I}_0(\lambda,x) = \int^{\infty}_{0} e^{-\lambda y}
  \Ei[\min(x,y)]dy \label{eq:72}
\end{equation}
and then differentiate with respect to $\lambda$.

For this we note that:
\begin{equation}
  \partial_\lambda (\Ei(x(1-\lambda))-\ln(1-\lambda)) =
  \frac{e^{x(1-\lambda)}}{\lambda-1} =
  (-x){_1F_1(1,2,x(1-\lambda))} \label{eq:73}
\end{equation}
and using the formula for derivatives of ${_1F_1}$, we get that for $i>0$:
\begin{equation}
  \partial_\lambda^{(i)} (\Ei(x(1-\lambda))-\log(1-\lambda)) =
  \frac{(-x)^i}{i} {_1F_1(i,i+1,x(1-\lambda))}. \label{eq:74}
\end{equation}
Finally, we can express:
\begin{align}
  \mathcal{I}_q(\lambda,x)
  &= (-1)^q\partial_\lambda^{(q)}
  I_0(\lambda,x) = (-1)^q \sum_{i=0}^q {{q}\choose{i}}
  \partial_\lambda^{(i)}(\Ei(x(1-\lambda)) - \log(1-\lambda))
  \partial_\lambda^{(q-i)} \lambda^{-1} \nonumber
  \\
  &= \frac{q!}{\lambda^{q}} I_0(\lambda,x) + \sum_{i=1}^q \frac{q!}{i!}
  \frac{x^i}{i}
  \frac{{_1F_1}[i,i+1,x(1-\lambda)]}{\lambda^{1+q-i}}. \label{eq:75}
\end{align}

The generalization for the case $\mu \neq 1$ is performed via a change
of variables
\begin{equation}
  \int^{\infty}_{0} {y}^q e^{-\lambda y} \Ei[\mu \min(x,y)] dy =
  \mu^{-1-q}
  \mathcal{I}_q\left(\frac{\lambda}{\mu},x \mu\right). \label{eq:76}
\end{equation}

As in the previous section, when $\mu = \lambda$ there is an
indeterminacy in the expression. In order to handle this case, we need
to take the limit of $\lambda \to 1$ in the above
expression~(\ref{eq:76}), to get:
\begin{equation}
  \mathcal{I}_q(1,x) = q! \mathcal{I}_0(1,x) + \sum_{i=1}^q
  \frac{q!}{i!} \frac{x^i}{i} = q!\left(\ln x + \gamma + \sum_{i=1}^q
  \frac{x^i}{i i!}\right). \label{eq:77}
\end{equation}

Finally we integrate \eqref{eq:56} to get:
\begin{align}
  &\int_0^\infty e^{-\lambda y} y^q G_{nl}^{(\mathrm{nsg})}(x,y)dy = \nonumber
  \\
  &\mspace{40mu} e^{\frac{-x}{n}} \sum_{i_1=l+1}^{n} \sum_{i_2=l+1}^{n}
    {{n+l}\choose{n-i_1}}
    \frac{(-1)^{i_1+i_2+1}}{(i_1-l-1)!(i_2-l-1)!}
    \left(\frac{2}{n}\right)^{i_1+i_2-2} \nonumber
  \\
  &\mspace{60mu}\times\Bigg[ \left(x^{i_1-1}
    \mathcal{F}_{q+i_2}(2/n-\alpha,x) + \frac{x^{i_2-1}}{\alpha^{i_1+q}}
    e^{\frac{x}{n}}\Gamma(i_1+q,x \alpha)\right) B_{i_2} \nonumber
  \\
  &\mspace{90mu}-x^{i_1-1} \frac{(q+i_2-1)!}{\alpha^{q+i_2}}
    \Bigg(\ln\left(\frac{4}{n^2}\frac{x}{\mu}\right) + \Psi(q+i_2) +
    \frac{2n+l-i_1+1}{i_1+l+1} \frac{x}{n^2} \nonumber
  \\
  &\mspace{280mu}+ \frac{2n+l-i_2+1}{i_2+l+1} \frac{i_2+q}{n^2 \alpha} +
    A_{i_1,i_2}\Bigg) \nonumber
  \\
  &\mspace{90mu}-x^{i_1-1}(n/2)^{q+i_2}\mathcal{I}_{q+i_2-1}(\alpha
    n/2,2x/n)\Bigg]. \label{eq:78}
\end{align}
For the case of $l \geq n$, we can immediately integrate the
expression for $G_{nl}(x,y)$, to get:
\begin{align}
  J_{nl}(\lambda, x)
  &= \int_0^{\infty} e^{-\lambda y} {y}^q
    G_{nl}(x,y)dy \nonumber
  \\
  &\mspace{40mu}=(-1)^{n+l}n \sum_{i=-l}^n \sum_{j=-l}^n
    \frac{(l-i)!}{(l+j)!}{{n+l}\choose{i+l}}
    \Big(\frac{2}{n}\Big)^{i+j} e^\frac{-x}{n} \nonumber
  \\
  &\mspace{60mu} \times \Bigg[{{l-j}\choose{l+n}} (-1)^{j+l}
    \left(x^{i-1}\mathcal{F}_{j+q}(2/n-\alpha,x) +
    \frac{x^{j-1}}{\alpha^{i+q}} e^{2/n x} \Gamma(i+q,x \alpha)\right)
    \nonumber
  \\
  &\mspace{90mu}-{{l-j}\choose{l-n}}
    \left(x^{i-1}\mathcal{F}_{j+q}(-\alpha,x) +\frac{x^{j-1}}{\alpha^{i+q}}
    \Gamma(i+q,x \alpha)\right) \Bigg]. \label{eq:79}
\end{align}

\subsection{Laurent series of $u_{a}^{b}$}
\label{sec:laurent-series-u_ab}

In order to integrate the Green function with $q,q' <l$ we will need
an expansion of $u_{a}^{b}$ functions in the Laurent series in the
neighbourhood of the singularities when $a = -m + \delta$ and
$a + b = -n + 2\delta$. In order to obtain the Laurent series it is
most convenient to start from the infinite series representation
\eqref{eq:49}, and separate singular terms. For this we split the
series into two parts, namely $i = 0..n-1$ and $n..\infty$. The latter
part is finite and is reduced to:
\begin{align}
  \tilde{u}_a^b(x,y,n)
  &= \sum_{i=n}^\infty \frac{\Gamma(a+b+i)}{(a+i)
  x^{a+b+i}} \frac{(-y)^i}{i!} \nonumber
  \\
  &= \frac{\Gamma(a+b+n)}{x^{a+b+n}} \Gamma(a+n) (-y)^{n}
    {_3\tilde{F}_2}
    \left(1,a+n,a+b+n,1+n,1+a+n,-\frac{y}{x}\right), \label{eq:80}
\end{align}
where ${_3\tilde{F}_2}(a_{1},a_{2},a_{3},b_{1},b_{2},z)$ is the
reduced generalized hypergeometric function. Therefore, the divergent
terms appear only in the first part.

We now introduce the following useful relations. The expansion of the
gamma function at positive integers gives:
\begin{align}
  \Gamma(\delta + n) = \Gamma(n)(1+\delta \Psi(n)) +
  O(\delta^2), \label{eq:81}
\end{align}
while for negative:
\begin{align}
  \Gamma(\delta - n) =
  \frac{(-1)^n}{n!} \left(\frac{1}{\delta} + \Psi(n+1) +
  \frac{\delta}{2} \left(\Psi(n+1)^2 - \Psi^{(1)}(n+1) +
  \frac{\pi^2}{3}\right)\right)
  + O(\delta^2). \label{eq:82}
\end{align}

Other relevant functions are
\begin{align}
  \frac{1}{x^{\delta+n}} = x^{-n} \left(1 - \delta\ln x + \delta^2
  \frac{\ln^{2}x}{2}\right) + O(\delta^3) \label{eq:83}
\end{align}
and
\begin{align}
  \frac{1}{x+\delta} = \frac{1}{x} \left(1-\frac{\delta}{x}\right) +
  O(\delta^2). \label{eq:84}
\end{align}

In addition, we use the following expansions
\begin{align}
  \mathcal{A}_n(x)
  &= \frac{\Gamma(n+2\delta)}{\delta x^{n+2\delta}} =
  \frac{\Gamma(n)}{x^{n}} \left(\frac{1}{\delta} + 2\Psi(n) - 2\ln{x}
  \right) + O(\delta), \label{eq:85}
  \\
  \mathcal{B}_m^n(x)
  &=\frac{\Gamma(-n+2\delta)}{(m+\delta)x^{-n+2\delta}} =
  \frac{(-x)^n}{n!m} \left(\frac{1}{2\delta} + \Psi(n+1) - \ln{x} -
  \frac{1}{2m}\right) + O(\delta), \label{eq:86}
  \\
  \mathcal{C}_{n}(x)
  &= \frac{\Gamma(-n+2\delta)}{\delta x^{-n+2\delta}} =
    \frac{(-x)^n}{n!} \Bigg(\frac{1}{2\delta^2} + \frac{\Psi(n+1) -
    \ln{x}}{\delta} \nonumber
  \\
  &+\Psi(n+1)^2 - \Psi^{(1)}(n+1) + \frac{\pi^2}{3} -
    2\ln(x)\Psi(n+1) + \ln^{2}{x}\Bigg) + O(\delta). \label{eq:87}
\end{align}

This allows us to find the Laurent series of $u_{a}^{b}$ in four
different cases depending on the signs of $a$ and $b$:
\begin{itemize}
\item When $a$ is a non-positive integer, we get exactly one divergent
  term (with index $i=-a$):
  \begin{align}
    u_{-n+\delta}^{b+\delta} (x,y) = \sum_{i=0}^{n-1}
    \frac{\Gamma(b-n+i)}{(i-n)x^{b}} \frac{(-y)^i}{i!} +
    \mathcal{A}_b(x) \frac{(-y)^{n}}{(n)!} +
    \tilde{u}_{-n}^b(x,y,n+1). \label{eq:88}
  \end{align}
\item When $a+b$ is a non-positive integer and $a$ is positive, the
  first $-a-b$ terms diverge:
  \begin{align}
    u_{\delta+a}^{\delta-a-n} (x,y) = \sum_{i=0}^n
    \mathcal{B}_{n-i}^{a+i}(x) \frac{(-y)^i}{i!} +
    \tilde{u}_{a}^{-a-n} (x,y,n+1). \label{eq:89}
  \end{align}
\item When $a+b$ is a non-positive integer and $b$ is positive, the
  first $-a-b$ temrs diverge as $\Gamma(-m)$ and independently the
  term with $i=-a$ diverges as $1/(a - a)$:
  \begin{align}
    u_{\delta-m-b}^{\delta+b} (x,y)
    &= \sum_{i=0}^m \mathcal{B}_{i-m-b}^{m-i}(x)\frac{(-y)^i}{i!} +
    \sum_{i=1+m}^{m+b-1} \frac{\Gamma(i-m)}{(i-m-b)x^{i-m}}
    \frac{(-y)^i}{i!} \nonumber
    \\
    &+ \mathcal{A}_b(x) \frac{y^{m+b}}{(m+b)!}
    +\tilde{u}_{-m-b}^{b}(x,y,m+b+1). \label{eq:90}
  \end{align}
\item When $a$ and $b$ are both non-positive integers, the first
  $-a-b$ terms diverge as $\Gamma(-m-n)$ with the $i=-a$ term
  diverging faster, as $\Gamma(-m-n)/(a-a)$:
  \begin{align}
    u_{\delta-m}^{\delta-n} (x,y) = \sum_{i=0,i \neq m}^{m+n}
    \mathcal{B}_{i-m}^{m+n-i}(x) \frac{(-y)^i}{i!} + \mathcal{C}_n(x)
    \frac{(-y)^{m}}{m!} + \tilde{u}_{-m}^{-n}(x,y,m+n+1). \label{eq:91}
  \end{align}
\end{itemize}

Lastly, examining \eqref{eq:57} it is clear that we also need to find
the Laurent series representation of the function $u_{a}^{\prime b}$
defined as
\begin{equation}
  {u'}_a^b(x,y)=\frac{\Gamma(a)}{x^a}\frac{\Gamma(b)}{y^b}. \label{eq:92}
\end{equation}
In the case of $a$ being a non-positive integer and $b>0$, we get:
\begin{equation}
  {u'}_{\delta-n}^{\delta+b}(x,y) = \frac{\Gamma(b)}{y^b}
  \frac{(-x)^n}{n!} \left(\frac{1}{\delta} + \Psi(1+n) + \Psi(b) -
    \log(x y)\right). \label{eq:93}
\end{equation}

\subsection{Laurent series of $K$}
\label{sec:laurent-series-K}

We first consider the case of $l \geq n$. Whenever $q+q'+i+j \leq 0$
we get $\mathcal{B}$ divergences, whenever $j+q \leq 0$ and
$i+q' \geq 0$ we get $\mathcal{A}$ divergences and for $j+q \leq 0$
and $i+q' \leq 0$, we get the mixed divergence
$\mathcal{C}$. Separating the sums over $i$ and $j$ in those three
cases we get:
\begin{align}
  &\int e^{-\lambda x - \lambda' y} G_{nl}(x,y) x^{q} y^{q'} dx
    dy \label{eq:94}
  \\
  &\mspace{60mu}=(-1)^{n+l}n \Bigg\{\sum_{i=-l}^{\min[n,l-q]} (l-i)!
    {{l+n}\choose{i+l}} \left(\frac{2}{n}\right)^{i} \alpha^{-q'}
    \mathcal{C}_{-i-q}(\alpha') h_{n,l}\left(q',-\frac{n
    \alpha}{2}\right) \nonumber
  \\
  &\mspace{180mu}+\sum_{i=-l}^{\min[n,l-q']} (l-i)! {{l+n}\choose{i+l}}
    \left(\frac{2}{n}\right)^{i} {\alpha'}^{-q}
    \mathcal{C}_{-i-q'}(\alpha) h_{n,l}\left(q,-\frac{n
    \alpha'}{2}\right) \nonumber
  \\
  &\mspace{180mu}+\sum_{i=-q}^{n} (l-i)! {{l+n}\choose{i+l}}
    \left(\frac{2}{n}\right)^{i} \mathcal{A}_{i+q}(\alpha') \alpha^{-q'}
    h_{n,l}\left(q',-\frac{n
    \alpha}{2}\right) \nonumber
  \\
  &\mspace{180mu}+\sum_{i=-q'}^{n} (l-i)! {{l+n}\choose{i+l}}
    \left(\frac{2}{n}\right)^{i} \mathcal{A}_{i+q'}(\alpha)
    {\alpha'}^{-q} h_{n,l}\left(q,-\frac{n
    \alpha'}{2}\right) \nonumber
  \\
  &\mspace{180mu} + \sum_{i=-l}^{\min[n,l-q-q']} (l-i)! {{l+n}\choose{i+l}}
    \left(\frac{2}{n}\right)^{i} \left[b_i(q,q',\alpha,\alpha') +
    b_i(q',q,\alpha',\alpha)\right]\Bigg\} + O(\delta^0), \nonumber
\end{align}
where
\begin{align}
  h_{n,l}(q,x) = \sum_{i=q}^{l} \frac{x^i}{(l-i)!(i-q)!}
  \left[{{l+i}\choose{l+n}} (-1)^{i+l} \left(1+\frac{1}{x}\right)^{i-q} -
  {{l+i}\choose{l-n}} \right] \label{eq:95}
\end{align}
and
\begin{align}
  b_i(q,q',\alpha,\alpha')
  &= \sum_{j=-l}^{-i-q-q'} \sum_{k=0,k \neq
  -j-q'}^{-j-i-q-q'}
  \frac{\mathcal{B}_{j+q'+k}^{-i-j-q-q'-k}(\alpha')}{(l+j)!}
  \left(\frac{2}{n}\right)^{j}\frac{1}{k!} \nonumber
  \\
  &\mspace{120mu}\times \left[{{l-j}\choose{l+n}} (-1)^{j+l}
    \left(\frac{2}{n}-\alpha\right)^{k} - {{l-j}\choose{l-n}}
    (-\alpha)^{k} \right]
    \nonumber
  \\
  &= \frac{(-\alpha')^{l-q-q'-i}}{\delta} d_{n,l}\left(l-q-q'-i, l-q',
    -\frac{n \alpha'}{2}, -\frac{n \alpha}{2} \right) +
    O(\delta^0), \label{eq:96}
\end{align}
where
\begin{align}
  d_{n,l}(p,r,x,y) = \sum_{i=0}^{p} \sum_{j=0,j \neq r-i}^{p-i}
  \frac{x^{-i-j}y^j}{(p-j-i)!(j+i-r)i!j!} \left[{{2l-i}\choose{l+n}}
  (-1)^{i} \left(\frac{1}{y} + 1\right)^{j} - {{2l-i}\choose{l-n}}
  \right]. \label{eq:97}
\end{align}
In section \ref{sec:proofs} we show that both $h_{n,l}(q,x)$ and
$d_{n,l}(p,r,x,y)$ vanish for every integer value of $0 \leq q \leq l$
and $0 \leq p \leq l$ respectively, proving that \eqref{nl} converges
in these cases.

For the case of $n > l$, we only need to consider
$G_{nl}^{(\mathrm{sg})}(x,y)$, as $G_{nl}^{(\mathrm{nsg})}(x,y)$ never
contains any singularities. In the case of $l=0$, we get:
\begin{align}
  &\int e^{-\lambda x - \lambda'
    y}G_{n0}^{(\mathrm{sg})}(x,y)x^{q+\delta}y^{q'+\delta} dx dy
    \nonumber
  \\
  &=\sum_{i_1=0}^{n-1}{{n}\choose{n-i_1}}\frac{(-1)^{i_1}}{(i_1)!}
    \left(\frac{2}{n}\right)^{i_1-1} \Bigg[u_{q+\delta}^{q'+i_1+1+\delta}
    \left(\alpha', \alpha - \frac{2}{n}\right) +
    u_{q'+\delta}^{q+i_1+1+\delta}\left(\alpha,\alpha' -
    \frac{2}{n}\right) \nonumber
  \\
  &\mspace{260mu}-{u'}_{q'+\delta}^{1+q+i_1+\delta}
    (\alpha',\alpha) -
    {u'}_{q+\delta}^{1+q'
    	+i_1+\delta}
    (\alpha,\alpha')\Bigg] \label{eq:98}
\end{align}
and by the use of Eq.~\eqref{eq:88}:
\begin{align}
  &\int e^{-\lambda x - \lambda' y} G_{n0}^{(\mathrm{sg})}(x,y) x^{q+\delta}
    y^{q'+\delta} dx dy \nonumber
  \\
  & =\sum_{i_1=0}^{n-1}{{n}\choose{n-i_1}}\frac{(-1)^{i_1}}{(i_1)!}
    \left(\frac{2}{n}\right)^{i_1-1} \Bigg[\mathcal{A}_{q'+i_1+1}(\alpha')
    \frac{(2/n-\alpha)^{-q}}{(-q)!} + \mathcal{A}_{q+i_1+1}(\alpha)
    \frac{(2/n-\alpha')^{-q'}}{(-q')!} \nonumber
  \\
  &\mspace{260mu}-\frac{(q+i_1)!}{\alpha^{1+q+i_1}}
    \frac{(2/n-\alpha')^{-q'}}{\delta(-q')!} -
    \frac{(2/n-\alpha)^{-q}}{\delta(-q)!}
    \frac{(q'+i_1)!}{{\alpha'}^{1+q'+i_1}} \Bigg] +
    O[\delta], \label{eq:99}
\end{align}
in which the divergent part vanishes by the definition of
$\mathcal{A}$.
		
For the case of $l>0$, we get:
 \begin{align}
   &\int  e^{-\lambda x - \lambda' y} G_{nl}^{(\mathrm{sg})}(x,y) x^{q+\delta}
     y^{q'+\delta} dx dy \nonumber
   \\
   &=\sum_{i_1=l}^{n-1}\sum_{i_2=1-l}^{1+l} {{n+l}\choose{n-i_1-1}}
     (-1)^{i_1+l} \frac{(i_2-1+l)!}{(i_1-l)!}
     \Big(\frac{2}{n}\Big)^{i_1-i_2} \nonumber
   \\
   &\mspace{40mu}\times \Bigg[{{n-i_2}\choose{n-l-1}}
     \left[u_{q-i_2+1+\delta}^{q'+i_1+1+\delta}
     \left(\alpha', \alpha - \frac{2}{n}\right) +
     u_{q'-i_2+1+\delta}^{q+i_1+1+\delta} \Big(\alpha, \alpha' -
     \frac{2}{n}\Big)\right] \nonumber
   \\
   &\mspace{60mu}-{{n+l}\choose{n-1+i_2}}
     \left({u'}_{q'-i_2+1+\delta}^{1+q+i_1+\delta}
     (\alpha',\alpha) +
     {u'}_{q-i_2+1+\delta}^{1+q'
     	+i_1+\delta}
     (\alpha,\alpha')\right)
     \Bigg] \label{eq:100}
 \end{align}
 and expanding the divergent part with the help of Eq.~\eqref{eq:88}:
 \begin{align}
   &\int  e^{-\lambda x - \lambda' y} G_{nl}^{(\mathrm{sg})}(x,y)
     x^{q+\delta} y^{q'+\delta} dx dy \label{eq:101}
   \\
   &\mspace{60mu}= \sum_{i_1=l}^{n-1}
     \sum_{i_2=1+q}^{1+l} {{n+l}\choose{n-i_1-1}} (-1)^{i_1+l}
     \frac{(i_2-1+l)!}{(i_1-l)!} \left(\frac{2}{n}\right)^{i_1-i_2}
     \nonumber
   \\
   &\mspace{90mu} \times\frac{1}{\delta} \Bigg[{{n-i_2}\choose{n-l-1}}
     \frac{(q'+i_1)!}{{\alpha'}^{q'+i_1+1}}
     \frac{(2/n-\alpha)^{i_2-q-1}}{(i_2-q-1)!} - {{n+l}\choose{n-1+i_2}}
     \frac{(q'+i_1)!}{{\alpha'}^{1+q'+i_1}}
     \frac{(-\alpha)^{i_2-q-1}}{(i_2-q-1)!}\Bigg] \nonumber
   \\
   &\mspace{60mu}+\sum_{i_1=l}^{n-1} \sum_{i_2=1+q'}^{1+l}
     {{n+l}\choose{n-i_1-1}} (-1)^{i_1+l} \frac{(i_2-1+l)!}{(i_1-l)!}
     \left(\frac{2}{n}\right)^{i_1-i_2} \nonumber
   \\
   &\mspace{90mu}\times\frac{1}{\delta} \Bigg[{{n-i_2}\choose{n-l-1}}
     \frac{(q+i_1)!}{{\alpha}^{q+i_1+1}}
     \frac{(2/n-\alpha')^{i_2-q'-1}}{(i_2-q'-1)!} -
     {{n+l}\choose{n-1+i_2}} \frac{(q+i_1)!}{{\alpha}^{1+q+i_1}}
     \frac{(-\alpha')^{i_2-q'-1}}{(i_2-q'-1)!}\Bigg] + O(\delta^0),
     \nonumber
 \end{align}
we can perform the sum over $i_1$ and isolate the sum over $i_2$ to get:
\begin{align}
  &\int e^{-\lambda x - \lambda' y} G_{nl}^{(\mathrm{sg})}(x,y)
    x^{q+\delta} y^{q'+\delta} dx dy \nonumber
  \\
  &\mspace{40mu}=\left(\frac{2}{n}\right)^{l} {{n+l}\choose{n-l-1}}
    \frac{1}{\delta} \Bigg(\frac{(q+l)!}{\alpha^{1+l+q}}
    {_2F_1\left(1+l-n,1+l+q,2l+2,\frac{2}{n \alpha}\right)}
    \frac{c_{n,l}(q',-\frac{2}{n\alpha'})}{{\alpha'}^{q'+1}} \label{eq:102}
  \\
  & \mspace{230mu}+\frac{(q'+l)!}{{\alpha'}^{1+l+q'}}
    {_2F_1\left(1+l-n,1+l+q',2l+2,-\frac{2}{n \alpha'}\right)}
    \frac{c_{n,l}(q,-\frac{2}{n\alpha})}{\alpha^{q+1}} \Bigg) +
    O(\delta^0),
    \nonumber
\end{align}
where
\begin{align}
  c_{n,l}(q,x) = \sum_{i=1+q}^{1+l} \frac{\Gamma(i+l)}{(i-q-1)!} x^{i}
  \left[{{n-i}\choose{l+1-i}} \left(1+\frac{1}{x}\right)^{i-q-1} -
  {{n+l}\choose{l+1-i}} \right]. \label{eq:103}
\end{align}

In section \ref{sec:proofs} we show that $c_{n,l}(q,x)$ vanishes for
every integer value of $0 \leq q \leq l$, provided that $n>l$, proving
that \eqref{eq:57} indeed converges.

We now define the functions $f_{a}^{b}$ to be the constant term, i.e.,
$\sim \delta^{0}$ of the Laurent series of $u_a^b$, given in equations
\eqref{eq:88}, \eqref{eq:89}, \eqref{eq:90} and \eqref{eq:91}, thus
arriving at our main result for integer values of $q$ and $q'$.

\subsection{Laurent series of J}
\label{sec:laurent-series-J}
For the case of non-positive integer, we can write the Laurent
series of $\mathcal{F}_{-n}$, as:
\begin{align}
  \mathcal{F}_{\delta-n}(x,y)
  &= \sum_{i=0}^{n-1}
    \frac{x^{i}}{i-n+\delta} \frac{y^{i-n+\delta}}{i!} +
    \frac{x^n}{\delta} \frac{y^\delta}{n!} + \sum_{i=n+1}^{\infty}
    \frac{x^{i}}{i-n+\delta} \frac{y^{i-n+\delta}}{i!} \nonumber
  \\
  &=\sum_{i=0}^{n-1} \frac{x^{i}}{i-n} \frac{y^{i-n}}{i!} +
    \frac{x^n}{n!} \left(\frac{1}{\delta}+\ln y\right) + \frac{y
    x^{1+n}}{(1+n)!} {_2F_2}(1,1,2,2-a,xy) + O(\delta), \label{eq:104}
\end{align}
which gives the Laurent series in the case of $n>l$, as:
\begin{align}
  &\int^{\infty}_{0}e^{-\lambda
    y}{y}^{q+\delta}G_{nl}^{(\mathrm{sg})}(x, y)dy \nonumber
  \\
  &= e^{\frac{-x}{n}} \sum_{i_1=l}^{n-1} \sum_{i_2=1+q}^{1+l}
     {{n+l_1}\choose{n-i_1-1}} (-1)^{i_1+l}
     \frac{(i_2-1+l)!}{(i_1-l)!} \left(\frac{2}{n}\right)^{i_1-i_2}
    \nonumber
  \\
  &\mspace{60mu}\times\Bigg[{{n-i_2}\choose{n-l-1}} x^{i_1}
    \mathcal{F}_{q+1-i_2+\delta}(\alpha-2/n,x) - {{n+l}\choose{n-1+i_2}}
    x^{i_1}\frac{1}{\alpha^{1+q-i_2+\delta}}
    \Gamma[q-i_2+1+\delta]\Bigg] + O(\delta^0) \nonumber
  \\
  &= e^{\frac{-x}{n}} \sum_{i_1=l}^{n-1} \sum_{i_2=1+q}^{1+l}
    {{n+l_1}\choose{n-i_1-1}} (-1)^{i_1+l} \frac{(i_2-1+l)!}{(i_1-l)!}
    \left(\frac{2}{n}\right)^{i_1-i_2} \nonumber
  \\
  &\mspace{60mu}\times \Bigg[{{n-i_2}\choose{n-l-1}} x^{i_1}
    \frac{(\alpha-2/n)^{i_2-1-q}}{(i_2-1-q)!} \frac{1}{\delta} -
    {{n+l}\choose{n-1+i_2}} \frac{x^{i_1}}{\alpha^{1+q-i_2}}
    \frac{(-1)^{q-i_2+1}}{(i_2-q-1)!}
    \frac{1}{\delta}\Bigg] + O(\delta^0) \nonumber
  \\
  &={{n+l}\choose{n-l-1}} \frac{n}{2x}
    M_{n,l+1/2}\left(\frac{2x}{n}\right) \alpha^{-1-q}
    c_{n,l}\left(q,-\frac{n \alpha}{2}\right)\frac{1}{\delta} +
    O(\delta^0), \label{eq:105}
\end{align}
where $M$ is the first Whittaker function.

On the other hand, the case of $l \geq n$ comes out as:
\begin{align}
  &\int^{\infty}_{0}e^{-\lambda y}{y}^{q+\delta}G_{nl}(x,
    y)dy \label{eq:106}
  \\
  &\mspace{60mu}=(-1)^{n+l}n \sum_{i=-l}^n \sum_{j=-l}^{-q}
    \frac{(l-i)!}{(l+j)!} {{n+l}\choose{i+l}}
    \Big(\frac{2}{n}\Big)^{i+j} e^\frac{-x}{n}
     \nonumber
  \\ 
  &\mspace{90mu} \times\Big[{{l-j}\choose{l+n}} (-1)^{j+l}
    x^{i-1}\mathcal{F}_{j+q+\delta}(2/n-\alpha,x) - {{l-j}\choose{l-n}}
    (x^{i-1}\mathcal{F}_{j+q+\delta}(-\alpha,x))\Big] + O(\delta^0)
    \nonumber
  \\
  &\mspace{60mu}=(-1)^{n+l}n \sum_{i=-l}^n \sum_{j=-l}^{-q}
    \frac{(l-i)!}{(l+j)!} {{n+l}\choose{i+l}}
    \left(\frac{2}{n}\right)^{i+j} e^\frac{-x}{n} x^{i-1} \nonumber
  \\ 
  &\mspace{90mu}\times\Bigg[{{l-j}\choose{l+n}} (-1)^{j+l}
    \frac{(2/n-\alpha)^{-j-q}}{(-j-q)!} \frac{1}{\delta} -
    {{l-j}\choose{l-n}} \frac{(-\alpha)^{-j-q}}{(-j-q)!}
    \frac{1}{\delta}\Big] + O(\delta^0) \nonumber
  \\
  &\mspace{60mu}=(-1)^{n+l} 2 (2l)! \left(\frac{n}{2x}\right)^{l+1}
    {_1F_1}\left[-n-l,-2l,\frac{2n}{x}\right] (-\alpha)^{-q}
    h_{n,l}\left(q,-\frac{n \alpha}{2}\right) \frac{1}{\delta} +
    O(\delta^0). \nonumber
\end{align}
Since $c_{n,l}(q,x)$ and $h_{n,l}(q,x)$ both vanish for every integer
value of $0 \leq q \leq l$, we can redefine $\mathcal{F}_q(x,y)$ to
the constant term of it's Laurent series completing the derivation of
$J$.

\section{Proofs}
\label{sec:proofs}
 
\subsection{$c_{n,l}=0$}
Here we show that $c_{n,l}$ vanishes for all integers $q \leq l$ in
the case of $n>l$, while $h_{n,l}$ vanishes for all $q \leq l$ in the
case $l \geq n$.
 
The function $c_{n,l}$ is given by a difference of two finite
series. Let us call them $c^{(1)}_{n,l}$ and $c^{(2)}_{n,l}$. Using
binomial theorem we get:
\begin{align} 
  c_{n,l}^{(1)}(q,x)
  &= \sum_{i=q+1}^{1+l} \frac{\Gamma(i+l)}{(i-q-1)!}
  x^{i} {{n-i}\choose{l+1-i}} \left(1+\frac{1}{x}\right)^{i-q-1}
  \nonumber
  \\
  &= \sum_{i=q+1}^{1+l} \sum_{k=0}^{i-q-1} \frac{\Gamma(i+l)}{(i-q-1)!}
  {{i-q-1}\choose{k}} x^{i-k} {{n-i}\choose{l+1-i}}. \label{eq:107}
\end{align}
Relabeling indices and using properties of binomials, this transforms
into:
\begin{align}
  c_{n,l}^{(1)}(q,x)
  &= \sum_{i=q+1}^{l+1} x^{i}
    \frac{\Gamma(i+l)}{(i-q-1)!} \sum_{k=0}^{1+l-i}
    {{k+l+i-1}\choose{l+i-1}} {{n-k-i}\choose{n-l-1}} \nonumber
  \\
  &= \sum_{i=q+1}^{l+1}
    x^{i} \frac{\Gamma(i+l)}{(i-q-1)!} {{n+l}\choose{l-i+1}} =
    c_{n,l}^{(2)}(q,x), \label{eq:108}
\end{align}
where in the last step we have used \cite{prudnikov_integrals_1986}:
\begin{align}
  \sum_{i=0}^{b-c}{{a+k}\choose{a}}{{b-k}\choose{c}} =
  {{a+b+1}\choose{b-c}}. \label{eq:109}
\end{align}

\subsection{$h_{n,l}=0$}
Proceeding along those same lines for $h_{n,l}$, we can similarly use
the binomial theorem to write:
 \begin{align}
   h_{n,l}^{(1)}(q,x) = \sum_{i=q}^{l}
   \frac{(-1)^{l+i}x^i}{(l-i)!(i-q)!} {{l+i}\choose{l+n}}
   \left(1+\frac{1}{x}\right)^{i-q} = \sum_{i=q}^{l} \sum_{k=0}^{i-q}
   \frac{(-1)^{l+i}x^{i-k}}{(l-i)!(i-q)!} {{l+i}\choose{l+n}}
   {{i-q}\choose{k}} \label{eq:110}
\end{align}
and redefine summation indices to arrive to:
  \begin{align}
    h_{n,l}^{(1)}(q,x) = \sum_{i=q}^{l} \sum_{k=0}^{l-i}
    \frac{(-1)^{l+i+k}x^{i}}{(l-k-i)!(k-q+i)!} {{l+k+i}\choose{l+n}}
    {{k-q+i}\choose{k}}. \label{eq:111}
\end{align}
Finally, using properties of binomials we get:
\begin{align}
  h_{n,l}^{(1)}(q,x)
  &= \sum_{i=q}^{l} \sum_{k=0}^{l-i}
  \frac{(-1)^{i+l+k}x^{i}}{(l-i)!(i-q)!} {{l+k+i}\choose{l+n}}
  {{l-i}\choose{k}} \nonumber
  \\
  &= \sum_{i=q}^{l} \frac{x^{i}}{(l-i)!(i-q)!}
  {{l+i}\choose{l-n}} = h^{(2)}_{n,l}(q,x), \label{eq:112}
\end{align}
where in the last step we have used the following identity valid
whenever $b<a$ \cite{prudnikov_integrals_1986}:
\begin{align}
  \sum_{k=0}^{b}{{a+k}\choose{b+c}}{{b}\choose{k}} (-1)^{b+k} =
  {{a}\choose{c}}. \label{eq:113}
\end{align}

\subsection{$d_{n,l}=0$}

Here we show that $d_{n,l}(p,r,x,y)$ vanishes for all integer values
of $0 \leq p \leq 2 l$ and $0<r<l$, assuming that $l \geq n$.

Firstly, $d_{n,l}$ is given as a difference of two series, that we
denote, as $d_{n,l}^{(1)}$ and $d_{n,l}^{(2)}$. Thus we have:
\begin{align}
  d_{n,l}^{(1)}(p,r,x,y) = \sum_{i=0}^{p} \sum_{j=0,j \neq r-i}^{p-i}
  \frac{x^{-i-j}y^j}{(p-i-j)!(j+i-r)i!j!} {{2l-i}\choose{l+n}}
  (-1)^{i} \left(\frac{1}{y} + 1\right)^{j}. \label{eq:114}
\end{align}
We can use the binomial theorem to transform it to:
\begin{align}
  d_{n,l}^{(1)}(p,r,x,y) = \sum_{i=0}^{p} \sum_{j=0,j \neq r-i}^{p-i}
  \sum_{k=0}^{j}  \frac{x^{-i-j}y^{j-r}}{(p-i-j)!(j+i-r)i!j!}
  {{2l-i}\choose{l+n}} (-1)^{i} {{j}\choose{k}}. \label{eq:115}
\end{align}
Redefinition of the summation variables, by $k \rightarrow j$,
$i \rightarrow k$ and $j \rightarrow i+j-k$ and changing the order of
sums yield:
\begin{align}
  d_{n,l}^{(1)}(p,r,x,y) = \sum_{i=0}^{p} \sum_{j=0,j \neq r-i}^{p-i}
  \sum_{k=0}^i \frac{x^{-i-j}y^j (-1)^{k}}{(p-i-j)!(j+i-r)(i+j-k)!k!}
  {{2l-k}\choose{l+n}}  {{i+j-k}\choose{j}}. \label{eq:116}
\end{align}
Using properties of binomials we can write it as:
\begin{align}
  d_{n,l}^{(1)}(p,r,x,y) = \sum_{i=0}^{p} \sum_{j=0,j \neq r-i}^{p-i}
  \sum_{k=0}^i \frac{x^{-i-j}y^j}{(p-i-j)!(j+i-r)i!j!}
  {{2l-k}\choose{l+n}} (-1)^{k} {{i}\choose{k}}. \label{eq:117}
\end{align}
Finally, we use:
\begin{align}
  \sum_{k=0}^{c} {{a-i}\choose{b}} {{c}\choose{i}}(-1)^{i} =
  {{a-c}\choose{a-b}}, \label{eq:118}
\end{align}
to arrive at:
\begin{align}
  d_{n,l}^{(1)}(p,r,x,y) = \sum_{i=0}^{p} \sum_{j=0,j \neq r-i}^{p-i}
  \frac{x^{-i-j}y^j}{(p-i-j)!(j+i-r)i!j!} {{2l-i}\choose{l-n}} =
  d_{n,l}^{(2)}(p,r,x,y). \label{eq:119}
\end{align}
completing the proof.

\section{Conclusion}
\label{sec:conclusion}

In our work we calculated the generating integral and integral moments
of the RCGF that appear during the calculation of the matrix elements
in second-order perturbation theory of multi-electron atoms and
ions. Our closed form analytical results allow one to effectively
compute the generating integrals and are expressed through elementary
and special functions.

In contrast to previous works, we followed the approach of the direct
integration of the RCGF with the corresponding powers of $r,r'$ and
exponentials. The main complications came from the fact, that when
$0 \leq q,q' < l+1$, the integrals from the individual terms of the
RCGF possess singularities that are explicitly cancelled only upon
summing all parts of the integrals of RCGF together. For this reason,
we employed the strategy of first to derive expressions valid for
non-integer values of $q$ and $q'$ and then find the Laurent series of
\eqref{eq:46} and \eqref{eq:47} around $q = m + \delta$,
$q' = m' + \delta$, where $m$ and $m'$ are non-negative
integers. After this we demonstrated that for $q,q' \geq 0$ the
divergent parts, that is the terms proportional to $\delta^{-1}$ and
$\delta^{-2}$ always vanish.

We validated the evaluation times of the generating integrals for
different input parameters and found out that the results via our
analytical expressions are typically several orders of magnitude
faster than the corresponding ones via direct numerical
integrations. Moreover, our analytical expressions provide correct
results even in cases where the direct numerical integration typically
fails to converge.

\begin{acknowledgments}            
  The authors are grateful to I. D. Feranchuk, C. H. Keitel,
  A. U. Leonau, S. Bragin, N. Oreshkina and Z. Harman for useful
  discussions. This article comprises parts of the PhD thesis work of
  Kamil Dzikowski to be submitted to the Heidelberg University,
  Germany.
\end{acknowledgments}

\bibliography{analytic_second}

\end{document}